\documentclass[final,3p,times]{elsarticle}

\usepackage{amsmath,hyperref}
\usepackage{lineno}

\usepackage{epstopdf}
\journal{}

\usepackage{lineno,hyperref}
\usepackage{graphicx}
\usepackage{dcolumn}
\usepackage{bm}
\usepackage{epsfig}
\usepackage{booktabs}
\usepackage{subfigure}
\usepackage{graphics}
\usepackage{amssymb}
\usepackage{amsmath}
\usepackage{array}
\usepackage{color}
\usepackage{booktabs}
\usepackage{multirow}
\usepackage{caption}
\usepackage{chngpage}
\usepackage{subfigure}
\usepackage{mathrsfs,gensymb,float}

\biboptions{numbers,sort&compress}
\modulolinenumbers[1]

\begin{document}

	\captionsetup[figure]{labelfont={bf},name={Fig.},labelsep=period}        
	
	\begin{frontmatter}
		
		\title{Phase-field based lattice Boltzmann method for containerless freezing}
		

		\author[1]{Jiangxu Huang}
		\author[2]{Lei Wang}
		\author[1,3,4]{Zhenhua Chai\corref{mycorrespondingauthor}}	
		\cortext[mycorrespondingauthor]{Corresponding author}
		\ead{hustczh@hust.edu.cn}
		\author[1,3,4]{Baochang Shi}

		\address[1]{ School of Mathematics and Statistics, Huazhong University of Science and Technology, Wuhan 430074, China}
		\address[2]{School of Mathematics and Physics, China University of Geosciences, Wuhan 430074, China}		
		\address[3]{Institute of Interdisciplinary Research for Mathematics and Applied Science,Huazhong University of Science and Technology, Wuhan 430074, China}	
		\address[4]{Hubei Key Laboratory of Engineering Modeling and Scientific Computing, Huazhong University of Science and Technology, Wuhan 430074, China}


\begin{abstract}
			
In this paper, a phase-field based lattice Boltzmann (LB) method is proposed to simulate solid-liquid phase change phenomena in multiphase systems. The method couples the thermal properties of the solidification front with the dynamics of the droplet interface, which enables the description of the complex interfacial changes during solid-liquid phase change process. The volume expansion or shrinkage of the droplet caused by the change of density during phase change is represented by adding a mass source term to the continuum equation. We first test the developed LB method by the three-phase Stefan problem and the droplet solidification on a cold surface, and the numerical results are in good agreement with the analytical and experimental solutions. Then the LB method is used to study the solidification problem with bubbles. The results show that the model can accurately capture the effect of bubbles on the solidification process, which is in good agreement with the previous work. Finally, a parametric study is carried out to examine the influences of some physical parameters on the sessile droplet solidification, and it is found that the time of droplet solidification increases with the increase of  droplet volume and contact angle.

\end{abstract}

		
\end{frontmatter}
	
	\section{Introduction}
	
	Solidification as a common solid-liquid phase change phenomenon is of significant importance in nature and industrial applications \cite{YaoAHT1989,BernsteinWF2000}, such as refrigeration \cite{LiuAE2012}, aerospace \cite{CaoAST2018}, additive manufacturing \cite{WeiSR2015} and food processing \cite{ArcherIJFM2004}. In the past decades, many theoretical and experimental studies have been conducted to investigate the fluid flows and heat transfer during solidification process. However, the solidification often occurs in an ambient fluid, where the phase change process and its interaction with the surrounding fluid need to be considered \cite{CaoACME2020,TiwariATE2023}. This process involves the gas, liquid and solid three phases, the dynamic evolution of the phase interfaces, as well as the coupling of flow and heat transfer ,bring some significant challenges to the investigation of solidification in multiphase systems.

	A single droplet freezing on the cold substrate has been widely considered as a fundamental case to explore the underlying freezing mechanism. Up to now some experimental works have been conducted to study the droplet freezing process on the cold substrate \cite{HuIJMF2010,ChaudharyETFS2014,MarinPRL2014,SchrembPRE2016,ZengPRF2022}. For instance, Hu et al. \cite{HuIJMF2010} experimentally investigated the droplet freezing on the cold substrate based on the molecular labeling thermometry technique. The results show that the volume of water droplet expands during freezing, and the expansion mainly in the upward direction rather than the radial direction. Chaudhary et al. \cite{ChaudharyETFS2014} experimentally studied the freezing of water droplet on a cooled surface. They used an infrared camera and thermocouple to measure the temperature evolution of the frozen droplet, observed four distinct processes: liquid cooling, recalescence, freezing, and solid cooling. Marin et al. \cite{MarinPRL2014} found that the liquid-solid interface of water droplet expands during freezing, causing the droplet to form pointed tip at the top, and the results show that this phenomenon is a consequence of a self-similar geometric mechanism, independent of the solidification rate, as evidenced by its lack of dependence on substrate temperature and wetting angle. Based on the work of Marin et al. \cite{MarinPRL2014}, Schremb et al. \cite{SchrembPRE2016} proposed a new experimental method to study the solidification of supercooled droplet using Hele-Shaw cell, and this method allows observation of the process in a quasi-two-dimensional manner, without optical distortions arising from the free surface of droplet. The experimental results show that the a small mutual influence of the dendrites is only observed when the freezing process is dominated by heat diffusion, and supercooling is higher. Recently, Zeng et al. \cite{ZengPRF2022} investigated the influence of gravity on the freezing of pendent and sessile droplet through experiments. They demonstrated that the gravity significantly affects droplet freezing process by shaping the initial droplet, resulting in the flattening or elongation of pendent and sessile droplet, respectively.

	Although experimental methods can be used to obtain external and localized internal information of frozen droplet through direct physical observation and measurement. However, due to the limitations of measurement techniques, it is difficult to obtain detailed information on internal icing front, temperature field, and velocity and pressure distributions inside the droplet. In particular,  although the intrusive measurement method can obtain information about the interior of the droplet, the measurement probe may affect the icing process. Therefore, in order to overcome the limitations of experimental methods, it is necessary to develop accurate theoretical models or numerical methods to derive more detailed information about the interior of frozen droplet. Through considering the effects of supercooling and gravity, Zhang et al. \cite{ZhangATE2017} developed a theoretical model to investigate the freezing behavior of water droplet, compared it with experiments. It is found that the freezing rate and time, as well as the droplet profile calculated by the model are in good agreement with the experimental results. Based on a one-dimensional approximation, Tembely et al. \cite{TembelyJFM2019} proposed a theoretical and numerical method to simulate droplet freezing on cold hydrophilic surfaces, and the model accurately predict the freezing time, the droplet volume expansion and tip singularity during freezing. Zhu et al. \cite{ZhuIJTS2022} conducted the experimental and theoretical studies on the freezing characteristics of water droplet deposited on the cold hydrophilic and hydrophobic aluminum surfaces. The results show that the freezing shape of sessile droplet depends on the surface temperature and wettability, and a power-law relationship between the freezing time of deposited droplet and the surface temperature was obtained. Although the theoretical models can be used to predict the freezing time, volume expansion and tip singularity that are compatible with experiments, they often rely on simplifying assumptions. For example, in the models of Zhang \cite{ZhangATE2017} and Zhu \cite{ZhuIJTS2022}, the freezing front is always assumed to be flat, which is clearly unreasonable; while in the theoretical model of Tembely et al.\cite{TembelyJFM2019},  the lubrication approximation is mainly applicable to the hydrophilic configurations. In addition, the dynamic evolution and strong nonlinear properties of phase interfaces also bring some challenges to the theoretical analysis.

	To overcome the limitations of the experimental and theoretical approaches, some numerical methods have been developed to study the solidification processes in multiphase systems \cite{VuIJMF2015,ShetabivashJCP2020,LyuJCP2021,SunIJR2015,ZhangPRE2020,XiongICHMT2018,XiongIJHMT2018,ZhangEnergy2023,MohammadipourJFM2024}. Vu et al. \cite{VuIJMF2015} and Shetabivash et al. \cite{ShetabivashJCP2020} simulated the solidification of sessile droplet on a cold plate by the front-tracking method and multiple level-set approach. In their works, the droplet volume change due to density variation during solidification are taken into account, and the angle at the tri-junction point can also be imposed. Lyu et al. \cite{LyuJCP2021} developed a novel hybrid volume-of-fluid and immersed-boundary (VOF-IB) method to simulate freezing droplet with considering the volume expansion during freezing, and test the accuracy of the method through a comparison with some available experimental and theoretical results. The numerical results in their work also show that the lower the density ratio, the longer the freezing time, and the more likely the formation of singular tips at the end of the freezing stage. Sun et al. \cite{SunIJR2015} and Zhang et al. \cite{ZhangPRE2020} investigated droplet solidification on the cold surface using the multiphase-field pseudopotential LB method, the main difference between the two model is that the volume change of the droplet during the freezing process is not taken into account in the work of Sun et al. \cite{SunIJR2015}. Xiong et al. \cite{XiongICHMT2018,XiongIJHMT2018} also employed the pseudopotential LB method to study the impact dynamics and solidification behavior of droplet on cold smooth and rough substrates. However, the pseudopotential LB method usually suffers form the numerical instability for the multiphase problems with large-density ratios \cite{ChenIJHMT2014}. To resolve this problem, the phase-field base LB method is adopted to study the freezing process of the droplet. For example, Zhang et al. \cite{ZhangEnergy2023} proposed a phase-field based LB method to study the ice evolution during methane hydrate dissociation, while volume change of droplet has not been considered. To overcome this drawback, Mohammadipour et al. \cite{MohammadipourJFM2024} developed another phase-field based LB method to investigate the solidification behavior of droplet in multicomponent systems, and considered the volume change of droplet during solidification process by adding a mass source term to the Cahn–Hilliard (CH) equation. However, it should be noted that in the framework of LB method, the fourth-order C-H equation cannot be recovered correctly \cite{WangCap2019}.

	In this work, we will propose a new phase-field based LB method to study containerless freezing problems, and simultaneously, the volume change during solidification is taken into account. Different from the previous work of Mohammadipour et al. \cite{MohammadipourJFM2024}, we considered a new second-order Allen-Cahn (A-C) equation where the volume change during solidification is included through adding a mass source term based on mass conservation. Also, the present LB method has more advantages in the study of multiphase flows with larger density ratios and in capturing topological changes of the interfaces. The remainder of this paper is organized as follows. In Sec. \ref{sec2}, mathematical model for the containerless freezing problems is proposed, followed by the developed phase-field LB method in Sec. \ref{sec3}. Numerical results and discussion are presented in Sec. \ref{sec4}, and finally, a brief summary is given in Sec. \ref{sec5}.

	\section{Mathematical model}
	\label{sec2}
\begin{figure}[H]
		\centering
		\includegraphics[width=0.5\textwidth]{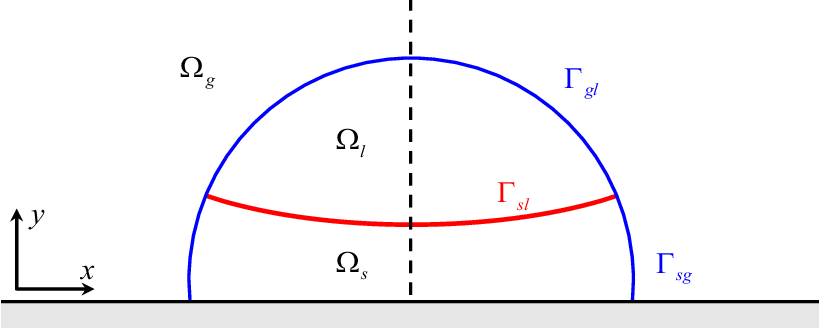}
		\caption{ Schematic of freezing droplet on a cold substrate.}
		\label{fig1}
	\end{figure}

	We now take the freezing process of a droplet on the cold substrate as an example, as shown in Fig. \ref{fig1}, and assume the solid phase only forms within the liquid phase, the liquid and gas phases are immiscible. The freezing process occurring in an ambient gas environment is a gas-liquid-solid ternary phase system, and the solid-liquid ($\Gamma_{s l}$), solid-gas ($\Gamma_{s g}$), and gas-liquid ($\Gamma_{g l}$) interfaces must be updated simultaneously. To simplify the following analysis, the phase interfaces can be divided into the one with phase change ($\Gamma_{s l}$) and those without phase change ($\Gamma_{s g}$, $\Gamma_{g l}$). The basic idea of present work is to use phase-field method to track interfaces without phase change and update the interface with phase change using enthalpy method. To this end, the phase-field order parameter, $\phi$, is applied to distinguish whether the medium is solid-liquid mixture or gas phase. Here $\phi = 1$ represents the solid-liquid mixtures while $\phi = 0$ denotes the surrounding gas phase. On the other hand, the solid fraction in the enthalpy method, $f_s$, is adopted to distinguish the interface between the solid and liquid phases in solid-liquid mixtures, $f_s=1$  stands for solid phase and $ f_s=0$ denotes liquid phase. With the help of order parameter $\phi$ and solid fraction $f_s$, the liquid, solid and gas phases can be represented by ($\phi=1$ $\&$ $f_s=0$) , ($\phi=1 $ $\&$ $ f_s=1$) and ($\phi=0 $ $\&$ $ f_s=0$). In this case, the physical properties of system can be characterized by a linear function of the order parameter and solid fraction: 
	\begin{equation}
		\zeta=f_s \zeta_s+\left(1-f_s\right) \phi \zeta_l+\left(1-f_s\right)(1-\phi) \zeta_g,
	\end{equation}
	where the parameter $\zeta$ denotes the density, viscosity, thermal conductivity and heat capacity, the subscripts $g$, $l$ and $s$ represents the gas, liquid and solid phases, respectively.
	
	In the following section, we will propose a mathematical model for containerless freezing processes, which includes the phase-field equation, enthalpy based energy equation, and Navier-Stokes equations.

	\subsection{Phase-field method for capturing interfaces without phase-change}
	The Allen-Cahn equation has been widely used to model moving interfaces among different phases \cite{ZhangEnergy2023,LiangPRE2018}, and for the interfacial dynamics of interfaces without phase-change ($\Gamma_{s g}$, $\Gamma_{g l}$), it can be written as
	\begin{equation}
		\frac{\partial \phi}{\partial t}+\nabla \cdot(\phi \mathbf{u})=\nabla \cdot[M(\nabla \phi-\lambda \mathbf{n})] + \phi \nabla \cdot\mathbf{u},
		\label{phasefield}
	\end{equation}
	where $\phi$ is the order parameter, $ \mathbf{u} $ is the velocity, $M$ is a positive constant named mobility, $ \mathbf{n}=\nabla \phi /|\nabla \phi|$ is the unit vector normal to the interface, $ \lambda $ is a function of $\phi$ and is defined as
	\begin{equation}
		\lambda=\frac{4 \phi(1-\phi)}{W},
	\end{equation}
	where $ W $ is the interface thickness. It should be noted that for incompressible fluid flows, the last term on the right hand side of Eq. (\ref{phasefield}) can be neglected, while for the freezing process including the volume change, it must be considered (see the following discussion).

	\subsection{Enthalpy method for evolving interface with phase- change}
	
	The temperature equation used to describe  the freezing front ($\Gamma_{s l}$) can be  derived from the energy conservation \cite{ChakrabortyJFM2007, HuangIJHMT2013}. 
	\begin{equation}
		\frac{\partial\left(\rho C_p T\right)}{\partial t}+\nabla \cdot\left(\rho C_p T \boldsymbol{u}\right)=\nabla \cdot(k \nabla T)+\dot{q},
		\label{eq4}
	\end{equation}
	where $ T $, $ C_p $ and $k$ are temperature, specific heat at constant pressure and thermal conductivity. $\dot{q}$ is the heat source term caused by the absorption or release of latent heat, and can be given by \cite{HuangIJHMT2013}
	\begin{equation}
		\dot{q}=-\left[\frac{\partial(\rho \Delta H)}{\partial t}+\nabla \cdot(\rho \boldsymbol{u} \Delta H)\right],
	\end{equation}
	where $ \Delta H $ is the latent enthalpy undergoing phase change. For pure material freezing, the second term $ \nabla \cdot(\rho \boldsymbol{u} \Delta H) $ can be ignored due to the uniform latent heat of the liquid, thus $\dot{q}$ can be simplified as \cite{HuangIJHMT2013}
	\begin{equation}
		\dot{q}=-\frac{\partial(\rho \Delta H)}{\partial t}=-\frac{\partial\left(\rho L f_l\right)}{\partial t},
		\label{eq6}
	\end{equation}
	where $ L $ is the latent heat, $f_l$ is the liquid fraction, given as $f_l = \Delta H / L$. Substituting Eq. (\ref{eq4}) into Eq. (\ref{eq6}), one can obtain \cite{ChakrabortyJFM2007, HuangIJHMT2013}
	\begin{equation}
		\frac{\partial \left(\rho H\right)}{\partial t}+\nabla \cdot\left(\rho C_p T \mathbf{u}\right)=\nabla \cdot(k \nabla T),
		\label{eq7}
	\end{equation}
	where $H$ is the total enthalpy, which can be divided into sensible and latent enthalpy components as
	\begin{equation}
		H=C_p T+L f_l.
		\label{eq8}
	\end{equation}
	Based on the definition of total enthalpy in Eq. (\ref{eq8}), the liquid fraction and temperature can be calculated as follows \cite{ShyyH1995}
	\begin{subequations}
		\begin{equation}
			f_l=\left\{\begin{array}{cc}
				0, & H \leq H_s \\
				\frac{H-H_s}{H_l-H_s}, & H_s \leq H \leq H_l \\
				1, & H \geq H_l
			\end{array}\right.
		\end{equation}	
		\begin{equation}
			T= \begin{cases}H / C_p & H<H_s \\ T_s+\frac{H-H_s}{H_l-H_s}\left(T_l-T_s\right) & H_s \leqslant H \leqslant H_l \\ T_l+\left(H-H_l\right) / C_p & H>H_l\end{cases}
		\end{equation}
	\end{subequations}
	where $T_s$ and $T_l$ are the solidus and liquidus temperatures, respectively, $H_s=C_{p, s} T_s$ and $H_s=C_{p, l} T_l+L$ are the total enthalpy at the solidus and liquidus temperatures. Through solving the enthalpy-based energy equation (\ref{eq7}), one can not only obtain the temperature field, but also simultaneously determine the liquid fraction, thereby achieving implicit tracking of the solid-liquid phase interface.

	\subsection{The Navier-Stokes equations for fluid flows}
	Apart from the interface-capturing equation mentioned above, we now introduce the governing equations for fluid flows. We assumed the fluid to be immiscible and Newtonian, and the fluid flows can be described by the following Navier–Stokes (N-S) equations \cite{JacqminJCP1999,LiangPRE2018}
	\begin{subequations} 
		\begin{equation}
			\nabla \cdot \mathbf{u}= 0,
		\end{equation}	
		\begin{equation}
			\frac{\partial \rho \mathbf{u}}{\partial t}+\nabla \cdot(\rho \mathbf{u u})=-\nabla p+\nabla \cdot\left[\mu\left(\nabla \mathbf{u}+(\nabla \mathbf{u})^{\mathrm{T}}\right)\right]+\mathbf{F}_s+\mathbf{G},
			\label{ns1}
		\end{equation}
	\end{subequations}
	where $ \rho $ is density, $ p $ is pressure, $ \mu $ is dynamic viscosity, $ \mathbf{G} $ is the body force, and $ \mathbf{F}_s $ is the surface tension force,
	\begin{equation}
		\mathbf{F}_s=\mu_\phi \nabla \phi,
		\label{eq10}
	\end{equation}
	where $ \mu_\phi $ is the chemical potential, and is defined by
	\begin{equation}
		\mu_\phi=4 \beta\left(\phi-\phi_l\right)\left(\phi-\phi_g\right)\left(\phi-\frac{\phi_l+\phi_g}{2}\right)-\kappa \nabla^2 \phi.
	\end{equation}
	The physical parameters $\beta$ and $ \kappa $ are related to the interface thickness $W$ and the surface tension $\sigma$,
	\begin{equation}
		k=\frac{3}{2} \sigma W, \quad \beta=\frac{12 \sigma}{W} .
	\end{equation}

	During the freezing process, the droplet volume may expand or shrink because of the density difference between solid and liquid phases \cite{HuIJMF2010,MarinPRL2014}. In order to include the volume change into the above N-S equations, some modifications have been made to the continuity equation and the tracer advection equation by neglecting the gas phase to account for the density change during the solidification of liquid-solid mixtures \cite{LyuJCP2021,ShetabivashJCP2020,RaessiNHTB2005}. Here we make a similar assumption adopted in the VOF and level-set frameworks through considering mixture of the three phases of liquid, solid, and gas. Actually, in an arbitrary control volume with the constant mass that consists of both solid and liquid phases, the conservation of mass can be expressed as
	\begin{equation}
		\frac{D}{D t}(M)=\frac{D}{D t}\left(M_l+M_s\right)=0
		\label{12}
	\end{equation}
	where the solid and liquid masses are defined as
	\begin{subequations}
		\begin{equation}
			M_s=\int_{V_{s}(t)} \rho_s d V_s=\int_{V(t)} \rho_s f_s d V,
			\label{13a}
		\end{equation}	
		\begin{equation}
			M_l=\int_{V_{l}(t)} \rho_l d V_l=\int_{V(t)} \rho_l (1-f_s) d V,
			\label{13b}
		\end{equation}
	\end{subequations}
	substituting Eqs. (\ref{13a}) and Eq. (\ref{13b}) in Eq. (\ref{12}), we can obtain
	\begin{equation}
		\frac{D}{D t}\left\{\int_{V(t)}\left[\rho_s f_s + \rho_l (1-f_s)\right] d V\right\}=0.
	\end{equation}
	Using Reynolds' transport theorem and assuming zero velocity in the solid phase, we have
	\begin{equation}
		\int_{V(t)} \left\{\frac{\partial}{\partial t} (\rho_s f_s) + \nabla \cdot [(1-f_s) \rho_l \mathbf{u}_l] - \frac{\partial}{\partial t} (\rho_l f_s)\right\} d V=0,
	\end{equation}
	where $\mathbf{u}_l$ is the liquid velocity. To ensure the integral over any integration region $V_{(t)}$ to be zero, one can obtain
	\begin{equation}
		\nabla \cdot [(1-f_s) \mathbf{u}_l]=(1-\frac{\rho_s}{\rho_l}) \frac{\partial f_s}{\partial t}.
	\end{equation}
	Using the fact $\mathbf{u}=\mathbf{u}_l (1-f_s)+\mathbf{u}_s f_s$, we can rewrite above equation as
	\begin{equation}
		\nabla \cdot \mathbf{u}=\dot{m},
		\label{ns2}
	\end{equation}
where source term $\dot{m}=(1-\frac{\rho_s}{\rho_l}) \frac{\partial f_s}{\partial t}$ on the right-hand side  of Eq. (\ref{ns2}) describes the expansion or shrinkage of the volume during the freezing process, and it is influenced by the ratio of solid density to liquid density.

In addition, how to treat the fluid-solid boundary is also a crucial issue. To overcome the difficulty in directly treating the solid-fluid interface, Noble and Torczynski \cite{NobleIJMPC1998} proposed an immersed moving boundary approach, which has also been widely used to deal with solid-liquid phase change interfaces \cite{ZhaoIJHMT2019, HeIJMF2023}. To more accurately characterize the interaction between fluid and solid, a diffuse-interface method is further developed through adding a modified external force term to the momentum equation \cite{LiuCF2022}. In this method, the momentum equation can be written as
\begin{equation}
	\frac{\partial \rho \mathbf{u}}{\partial t}+\nabla \cdot(\rho \mathbf{u u})=-\nabla p+\nabla \cdot\left[\mu\left(\nabla \mathbf{u}+(\nabla \mathbf{u})^{\mathrm{T}}\right)\right]+\mathbf{F}_s+\mathbf{G} +  \rho \mathbf{f},
	\label{ns3}
\end{equation}
where $\mathbf{f}$ is the force generated by fluid solid interaction. In summary, the mathematical model including equations (\ref{phasefield}), (\ref{eq7}), (\ref{ns2}), and (\ref{ns3}) is used to describe the freezing process of droplet considering volume change.

\section{LB method for containerless freezing}
\label{sec3}
In this section, we will develop a new LB method where three different LB models are adopted for phase field, temperature field and flow field.

\subsection{LB model for the phase field}
The evolution equation of LB model with the BGK collision operator for the Allen-Cahn equation can be written as \cite{LiangPRE2018} 
\begin{equation}
	g_i\left(\mathbf{x}+\mathbf{c}_i \Delta t, t+\Delta t\right)-g_i(\mathbf{x}, t)=-\frac{1}{\tau_g}\left[g_i(\mathbf{x}, t)-g_i^{\mathrm{eq}}(\mathbf{x}, t)\right]+ \left(1-\frac{1}{2 \tau_g}\right) \Delta t G_i(\mathbf{x}, t),
	\label{eq19}
\end{equation}
where $g_i(\mathbf{x}, t)$ is the order parameter distribution function at position $\mathbf{x}$ and time $t$, $\mathbf{c}_i$ is the discrete velocity. For the D2Q9 model considered here, the weight coefficient $\omega_i$ and discrete velocity $\mathbf{c}_i$ are defined as
\begin{equation}
	\omega_i= \begin{cases}4 / 9 & i=0, \\ 1 / 9 & i=1-4, \\ 1 / 36 & i=5-8,\end{cases},
\end{equation}
\begin{equation}
	\mathbf{c}_i= \begin{cases}(0,0), & i=0, \\ (\cos [(i-1) \pi / 2], \sin [(i-1) \pi / 2]) c, & i=1-4, \\ (\cos [(2 i-9) \pi / 4], \sin [(2 i-9) \pi / 4]) \sqrt{2} c, & i=5-8,\end{cases}
\end{equation}
where $c = \delta x/\Delta t$ is the lattice speed with $\delta x$ and $\delta t$ denoting the lattice spacing and time step, respectively (both of them are set to 1 in the present work). $g_i^{\mathrm{eq}}$ is the equilibrium distribution, and is given by
\begin{equation}
	g_i^{\mathrm{eq}}=\omega_i \phi\left(1+\frac{\mathbf{c}_i \cdot \mathbf{u}}{c_s^2}\right),
\end{equation}
where $c_s = c /\sqrt 3$ is the sound speed. To recover the Allen-Cahn equation with the multiscale analysis, the source term $G_i$ should be designed as 
\begin{equation}
	G_i= \frac{\omega_i \mathbf{c}_i \cdot\left[\partial_t(\phi \mathbf{u})+c_s^2 \lambda \mathbf{n}\right]}{c_s^2} + \omega_i \phi \nabla \cdot\mathbf{u}.
	\label{eq23}
\end{equation}
In addition, the order parameter in the present LB model can be computed by
\begin{equation}
	\phi=\sum_i g_i + \frac{\Delta t}{2} \phi \nabla \cdot\mathbf{u}.
\end{equation}
Following the Chapman-Enskog analysis, the Allen-Cahn equation can be recovered correctly from the LB equation (\ref{eq19}) with the mobility $M=c_s^2\left(\tau_f-0.5\right) \Delta t$.

\subsection{LB model for the temperature field} 

For the temperature field, the enthalpy-based thermal LB model proposed by Huang et al. \cite{HuangIJHMT2013} is adopted, and the evolution equation of the total enthalpy distribution function $h_i(\mathbf{x}, t)$ reads 
\begin{equation}
	h_i\left(\mathbf{x}+\mathbf{e}_i \Delta t, t+\Delta t\right)=h_i(\mathbf{x}, t)-\frac{1}{\tau_h}\left[h_i(\mathbf{x}, t)-h_i^{e q}(\mathbf{x}, t)\right],
\end{equation}
where $\tau_h$ is the relaxation time related to the thermal conductivity, $h_i^{e q}$ is the local equilibrium distribution function defined as \cite{HuangIJHMT2013}
\begin{equation}
	h_i^{e q} = \begin{cases}H-C_{p, \text { ref }} T+\omega_i C_p T\left(\frac{C_{p, \mathrm{ref}}}{C_p}-\frac{\mathbf{I}: \mathbf{u} \mathbf{u}}{2 c_s^2}\right), & i=0, \\ \omega_i C_p T\left[\frac{C_{p, \text { ref }}}{C_p}+\frac{\mathbf{e}_i \cdot \mathbf{u}}{c_s^2}+\frac{\left(\mathbf{e}_i \mathbf{e}_i-c_s^2 \mathbf{I}\right): \mathbf{u} \mathbf{u}}{2 c_s^4}\right], & i \neq 0,\end{cases}
\end{equation}
The total enthalpy is is calculated by
\begin{equation}
	H=\sum_{i=0} h_i.
\end{equation}

\subsection{LB model for the flow field}
Different from the original LB model for the flow field, we incorporate an extra mass source arising from density change during the freezing process, which results in some modifications to both the evolution equations and the computation of macroscopic quantities. The evolution equation of the LB model for flow field is formulated as follows \cite{YuanCMA2020}:
\begin{equation}
	\begin{aligned}
		f_i\left(\mathbf{x}+\mathbf{c}_i \Delta t, t+\Delta t\right)-f_i(\mathbf{x}, t)=  -\frac{1}{\tau_f}\left[f_i(\mathbf{x}, t)-f_i^{\mathrm{eq}}(\mathbf{x}, t)\right] +\Delta t \left(1-\frac{1}{2 \tau_f}\right)   F_i(\mathbf{x}, t),
	\end{aligned}
	\label{eqf}
\end{equation}
where $\tau_f = \nu / {c_s}^2 \Delta t +0.5$ is the corresponding relaxation time for flow field, $f_i^{\mathrm{eq}}$ is the local equilibrium distribution function,
\begin{equation}
	\begin{aligned}
		&f_i^{\mathrm{eq}}= \begin{cases}\frac{p}{c_s^2}\left(\omega_i-1\right)+\rho s_i(\mathbf{u}), & \mathrm{i}=0 \\ \frac{p}{c_s^2} \omega_i+\rho s_i(\mathbf{u}), & \mathrm{i} \neq 0\end{cases}
	\end{aligned}
\end{equation}
with
\begin{equation}
	\begin{aligned}
		s_i(\mathbf{u})=\omega_i\left[\frac{\mathbf{c}_i \cdot \mathbf{u}}{c_s^2}+\frac{\left(\mathbf{c}_i \cdot \mathbf{u}\right)^2}{2 c_s^4}-\frac{\mathbf{u} \cdot \mathbf{u}}{2 c_s^2}\right] .
	\end{aligned}
\end{equation}
Unlike from the previous LB models, here the forcing term $F_i$ is given by \cite{YuanCMA2020}
\begin{equation}
	F_i=\omega_i  \left[S +\frac{\mathbf{c}_i \cdot (\mathbf{F}+\rho \mathbf{f})}{c_s^2} + \frac{(\mathbf{u}\tilde{\mathbf{F}} + \mathbf{u}\tilde{\mathbf{F}}):\left(\mathbf{c}_i \mathbf{c}_i-c_s^2 \mathbf{I}\right)}{2c_s^4}\right],
\end{equation}
where $ S=\rho \dot{m} + \mathbf{u} \cdot \nabla \rho $, $ \tilde{\mathbf{F}} = \mathbf{F} - \nabla p + c_s^2 \nabla \rho +  c_s^2 \nabla \cdot S $, $\mathbf{F}=\mathbf{F}_{\mathrm{s}}+\mathbf{G}$ is the total force. The macroscopic velocity  $\mathbf{u}$ and  pressure $p$ can be evaluated by
\begin{equation}
	\rho \mathbf{u}^*=\sum \mathbf{c}_i f_i+\frac{\Delta t}{2}  \mathbf{F},
\end{equation}
\begin{equation}
	\mathbf{u}=\mathbf{u}^*+\frac{\Delta t}{2} \mathbf{f},
\end{equation}
\begin{equation}
	p=\frac{c_s^2}{\left(1-\omega_0\right)}\left[\sum_{i \neq 0} f_i+\frac{\Delta t}{2}S + \tau \Delta t F_0 + \rho s_0(\mathbf{u})\right],
\end{equation}
where $\mathbf{u}^*$ is the velocity without considering the fluid-particle interaction, $\mathbf{u}$ is the corrected velocity. The fluid-solid interaction $\mathbf{f}$ can be discretized as $f_s \left(\mathbf{u}_s-\mathbf{u}^*\right) / \Delta t$ \cite{LiuCF2022}, $\mathbf{u}_s$ is the solid-phase velocity. We would like to point out that this treatment has been successfully applied to deal with a variety of fluid-solid coupling problems, such as the particulate flows \cite{LiuCF2022}, dendrite growth \cite{ZhanARXIV2024} and two-phase flow in complex structures \cite{ZhanPD2024}.

In numerical simulations, the derivative terms should be discretized with some suitable difference schemes. For simplicity, the temporal derivative in Eq. (\ref{eq23}) is approximated by the first-order explicit Euler scheme \cite{LiuPRE2023} 
\begin{equation}
	\partial_t(\phi \mathbf{u})(\mathbf{x}, t)=[(\phi \mathbf{u})(\mathbf{x}, t)-(\phi \mathbf{u})(\mathbf{x}, t-\Delta t)] / \Delta t .
\end{equation}
To calculate the gradient and Laplace operator appeared above, the second-order isotropic central schemes are applied \cite{LiuPRE2023, LiangPRE2018}
\begin{subequations}
	\begin{equation}
		\nabla \zeta(\mathbf{x}, t)  =\sum_{i \neq 0} \frac{\omega_i \mathbf{c}_i \zeta\left(\mathbf{x}+\mathbf{c}_i \Delta t, t\right)}{c_s^2 \Delta t},
		\label{eqtidi1}
	\end{equation}	
	\begin{equation}
		\nabla^2 \zeta(\mathbf{x}, t)  =\sum_{i \neq 0} \frac{2 \omega_i \left[\zeta\left(\mathbf{x}+\mathbf{c}_i \Delta t, t\right)-\zeta(\mathbf{x}, t)\right]}{c_s^2 \Delta t^2}.
		\label{eqtidi2}
	\end{equation}
\end{subequations}

\subsection{Numerical implementation of the wetting boundary condition}

In the presence of gas-liquid-solid interaction, the fluid interface dynamics would be greatly affected by the wettability of the solid substrate, and consequently influencing the freezing behavior. Therefore, it is important to establish wetting boundary conditions to describe the interaction between the fluid and solid substrate. According to the geometric relationship, the wetting boundary condition can be formulated as \cite{DingPRE2007}:
\begin{equation}
	\mathbf{n}_w \cdot \nabla \phi=-\tan \left(\frac{\pi}{2}-\theta\right)\left|\mathbf{n}_\tau \cdot \nabla \phi\right|,
\end{equation}
where $\mathbf{n}_w$ represents the unit normal vector points from the solid wall into the fluids, $\mathbf{n}_\tau$ is the unit vector tangential to solid surface. $\mathbf{n}_w \cdot \nabla \phi$ and $\nabla \phi$, $\mathbf{n}_\tau \cdot \nabla \phi$ are the normal and tangential components of $\nabla \phi$, and they can be determined by the second-order schemes \cite{LiangPRE2019}
\begin{subequations}
	\begin{equation}
		n_{\mathrm{w}} \cdot \nabla \phi=\frac{\phi_{x, 1}-\phi_{x, 0}}{\Delta x},
	\end{equation}

	\begin{equation}
		n_\tau \cdot \nabla \phi=\frac{\partial \phi_{x, 1 / 2}}{\partial x}=1.5 \frac{\partial \phi_{x, 	1}}{\partial x}-0.5 \frac{\partial \phi_{x, 2}}{\partial x},
	\end{equation}	
\end{subequations}

where the lattice nodes near the solid substrate are divided into the fluid layer, solid boundary ($y = 1/2$) and ghost layer ($y = 0$). The derivatives of the order parameters can be calculated by a second-order central difference scheme \cite{LiangPRE2019}
\begin{equation}
	\frac{\partial \phi_{x, y}}{\partial x}=\frac{\partial \phi_{x+1, y}-\partial \phi_{x-1, y}}{2 \partial x} .
\end{equation}
Once the values of the order parameters at the ghost layer are obtained, the gradient of order parameter and the Laplacian operator at all fluid nodes can be calculated from Eqs. (\ref{eqtidi1}) and (\ref{eqtidi2}). 


\subsection{Dimensionless numbers}
According to the similarity theory, the numerical solution should be similar to the experimental results as long as they have geometric similarity and the same dimensionless parameters. For the multiphase solidification considered in this work, the system is mainly  governed by the following non-dimensional parameters:
\begin{equation}
	 \mathrm{Ste}=\frac{c_p\left(T_m-T_w\right)}{L}, \quad  \operatorname{Pr}=\frac{\mu_l c_p}{\lambda_l},\quad  \mathrm{Fo}=\frac{\lambda_l t}{\rho_l c_p R_0^2},
\end{equation}
where $\mathrm{Ste}$, $\mathrm{Pr}$, $\mathrm{Fo}$ are the Stefan number, Prandtl number and Fourier number.

\section{Numerical results and discussion}
\label{sec4}

In this section, we will test the capability and reliability of the proposed LB model with several typical benchmark problems, including the three-phase Stefan problem, freezing droplet on cold surface and rising bubbles with solidification, and conduct some comparisons between the present results with the available numerical, analytical, and experimental data reported in the previous works.

\subsection{Three-phase Stefan problem}
\begin{figure}[H]
	\centering
	\includegraphics[width=0.45\textwidth]{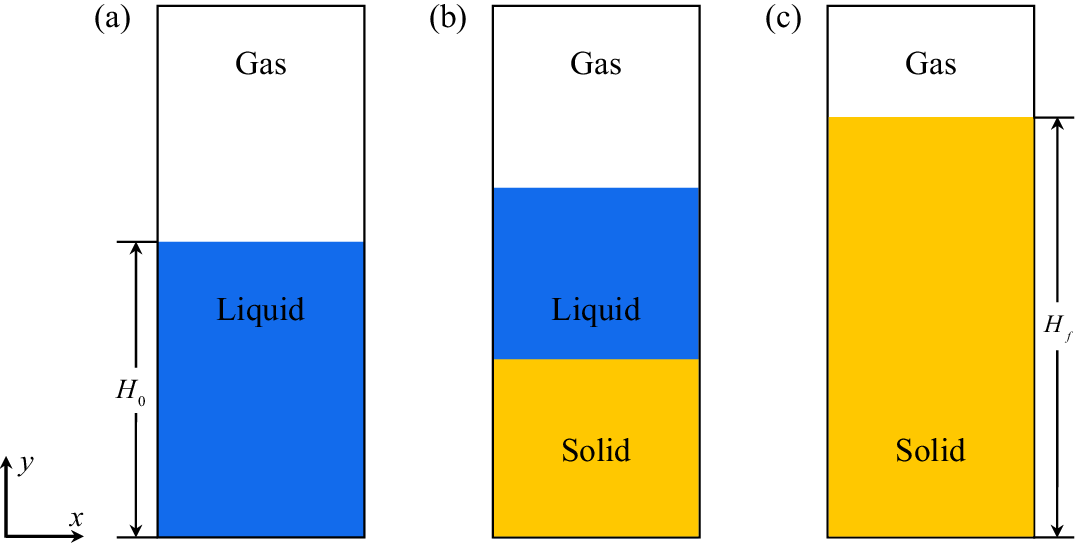}
	\caption{ Schematic of the three-phase Stefan problem. [The initial stage before freezing (a), the intermediate stage of partial liquid phase freezing to solid phase (b), the final moment of complete freezing (c)].}
	\label{fig2}
\end{figure}
In order to accurately predict the droplet freezing process, the volume change during the freezing process should be considered, to this end, a source term is added in the continuum equation, as described previously. We now focus on simulate the freezing of pure matter in the semi-infinite space to verify the applicability and accuracy of the method in treating volume change. The schematic of the problem is depicted in Fig. (\ref{fig2}). Initially, the liquid phase and gas phase are uniformly distributed in the regions $0 \leq y \leq H_0$ and  $H_0 \leq y \leq L$ with the uniform temperature $T_0$ ($T_0>T_m$, $T_m$ is the freezing temperature), then a constant temperature $T_w \left(T_w<T_m\right)$ is imposed on the bottom wall. To match this setup, the initial profile of the order parameter is given by
\begin{equation}
	\phi(x, y)=0.5+0.5 \tanh \frac{2\left(H_0-y\right)}{W},
	\label{stefan}
\end{equation}
where the interface width W is set as 5. With the evolution of freezing front from the bottom surface to the free surface, the interfaces among the liquid, gas, and solid phases should be treated simultaneously. Once the freezing of the liquid phase is completed, the height of the solid phase should be equal to the maximum value $H_f$. According to the conservation of mass, the final height of the frozen liquid can be determined from the following equation \cite{ShetabivashJCP2020}
\begin{equation}
	H_f=\frac{\rho_l}{\rho_s} H_0.
	\label{StefanP}
\end{equation} 

\begin{figure}[H]
	\centering
	\includegraphics[width=0.43\textwidth]{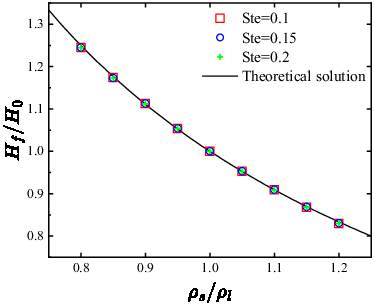}
	\caption{A Comparisons of final dimensionless solid-phase height $H_f/H_0$ between numerical results and theoretical solution (\ref{StefanP}) at different values of  $\rho_s/\rho_l$ after complete solidification.}
	\label{fig3}
\end{figure}

In our simulations, the grid resolution of the computational region is set to be $N_x \times N_y=400 \times 10$, the physical parameters are given as $T_0 = 0.1$, $T_w=-2$, $ T_m=0 $, $\mathrm{Ste}=0.1-0.2$, $C_{p,s}/C_{p,l} = 1$ and $\lambda_s / \lambda_l=1$. For the phase,  temperature  and flow fields, the bottom and top surfaces are the solid walls imposed by the no-flux, Dirichlet and no-slip boundary condition, while the periodic boundary condition is applied in the horizontal direction. To treat the no-flux and no-slip boundary conditions, the half-way bounce-bounce scheme is used, while for the Dirichlet boundary condition, the general bounce-back scheme \cite{ZhangPRE2012} is adopted. Fig. \ref{fig3} shows a comparison of $H_f/H_0$ at different values of  $\rho_s/\rho_l$ between the numerical results and theoretical solution (\ref{StefanP}), and a good agreement between them can be observed. This indicates that the present LB method can accurately capture the volume change during the freezing process and preserve the mass conservation.

\subsection{Droplet freezing on a cold surfaces}

Droplet freezing on the cold surface is a fundamental heat transfer problem that has been widely used to assess the proposed numerical methods for simulating freezing process. In this part, to show the capacity of the present LB method, we first simulate the droplet freezing on a cold substrate, and then we explore the effects of solid-liquid density ratio and wettability on the droplet freezing.

\subsubsection{A comparison between the numerical and experimental results}

\begin{figure}[H]
	\centering
	\includegraphics[width=0.7\textwidth]{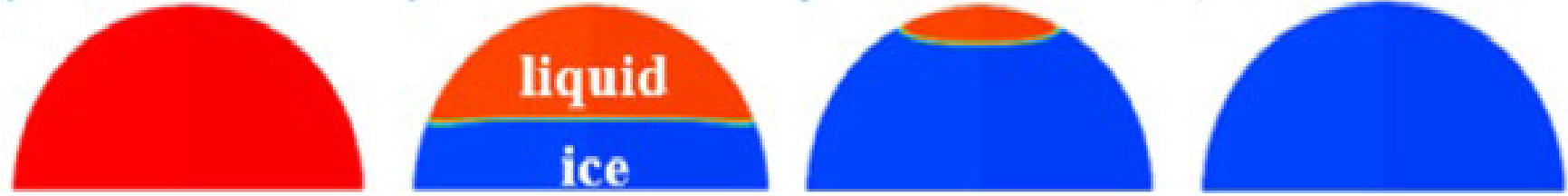} 
	\label{fig4a}
	\put(-350 ,35){\small (\textit{a})}
	\vspace{0.3cm}

	\includegraphics[width=0.7\textwidth]{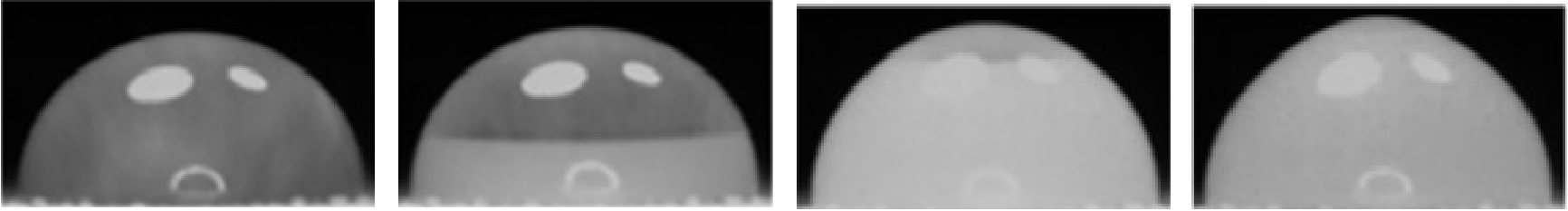} 
	\label{fig4b}
	\put(-350,35){\small (\textit{b})}
	\vspace{0.3cm}

	\includegraphics[width=0.7\textwidth]{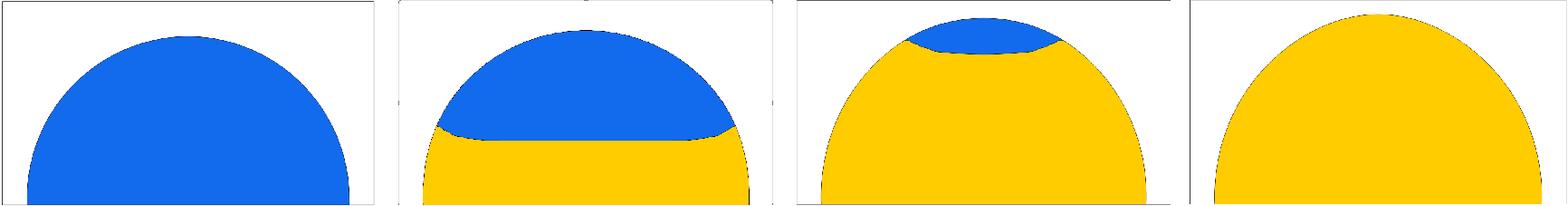} 
	\label{fig4c}	
	\put(-350,35){\small (\textit{c})}
	
	\caption{A comparison of the freezing shapes of droplet at different times [Ref. \cite{HouIJHMT2022} (a), experiment (b) and present LB model (c) at different times, where the gas, liquid, and solid phases are labeled by the white, blue and orange colors].}
	\label{fig4}
\end{figure}

We first conduct a comparison of numerical results with the experimental data reported in Ref. \cite{HouIJHMT2022}. In the experiment conducted by Hou et al. \cite{HouIJHMT2022}, a droplet with a volume of $15 \mu \mathrm{L}$ is gently deposited on a super-cooled surface at a temperature $ T=-29.5^{\circ} \mathrm{C}$ with the contact angle $\theta=86.4^{\circ}$. Our simulations are carried out in a two-dimensional domain with the grid resolution $N_x \times N_y=400 \times 200$, and the temperature is set to be $T_0$. Initially, a semicircular droplet with the radius of $R = 40$ is located at the center of the cold substrate. A lower temperature $T_w$ is imposed on the bottom surface after the contact angle of the droplet being equal to the prescribed value $\theta$. The distribution of the order parameter is initialized by
\begin{equation}
	\phi(x, y)=0.5+0.5 \tanh \frac{2\left[R-(x-x_0)^2-(y-y_0)^2\right]}{W},
\end{equation}
where $(x_0,y_0) = (0, N_x/2)$ is the coordinates of droplet center, $W=5$ is the interface thickness. The periodic boundary conditions applied in the horizontal direction, while the no-slip boundary condition is adopted at the bottom and top boundaries. Based on experimental conditions and the physical properties of water listed  in Table \ref{table1}, the following none-dimensional parameters, $\mathrm{Pr}=7.25$, $\mathrm{Ste}=0.02$, $k_{s}/k_{l}=3.8$, $C_{p,s}/C_{p,l}=0.5$, $\rho_{s}/ \rho_{l}=0.9$ and the contact angle $\theta=86.4^{\circ}$ are used.

\begin{table}[H]
	\centering
	\caption{ Physical properties of water and ice at $0^{\circ} \mathrm{C}$.}
	\begin{tabular}{ccccc}
		\hline \hline
		Material & \begin{tabular}[c]{@{}c@{}}Density $\rho$\\ $\left(\mathrm{kg}\ \mathrm{m}^{-3}\right)$\end{tabular} & \begin{tabular}[c]{@{}c@{}}Heat capacity $C_p$\\ $\left(\mathrm{kJ}\ \mathrm{kg}^{-1}\ \mathrm{~K}^{-1}\right)$\end{tabular} & \begin{tabular}[c]{@{}c@{}}Thermal conductivity $k$\\ $\left(\mathrm{W}\ \mathrm{m}^{-1}\ \mathrm{~K}^{-1}\right)$\end{tabular} & \begin{tabular}[c]{@{}c@{}}Latent heat $L$\\ $\left(\mathrm{kJ}\ \mathrm{kg}^{-1}\right)$\end{tabular} \\ \hline
		Water    & 999.8                                                                                                & 4.22                                                                                                                         & 0.56                                                                                                                            & 333.4                                                                                                  \\
		Ice      & 917.0                                                                                                & 2.02                                                                                                                         & 2.26                                                                                                                            & -                                                                                                       \\ \hline \hline
	\end{tabular}
	\label{table1}
\end{table}

Depending on the temperature fluctuation of water droplet, the freezing process of the droplet on a cold plate can be divided into five stages: liquid stage, nucleation, recalescence stage, solidification stage and post-solidification stage \cite{TiwariATE2023}. Due to the much shorter duration of the nucleation and recalescence stages, compared to the entire freezing period, it is difficult to simulate such a small timescale. Thus, we only focus on the freezing stage based on heat balance, and take the time before nucleation as the initial state \cite{ZhangPRE2020}. Fig. \ref{fig4} presents a comparison of the freezing process of water droplet on the cold surface between present results, previous experimental and numerical solutions. In the experiments, the freezing front starts to move from the bottom to the top of the droplet, the droplet volume expands continuously due to the density difference between water and ice, and the droplet shape expands mainly in the vertical direction instead of the radial direction. It is clear that the present results agree well with the experimental data, and the evolution of the melting front and the change of droplet shape can be captured accurately, which indicates that the present LB method can provide accurate numerical results in the study of the droplet freezing process. However, the numerical results without considering volume change during freezing process are inconsistent with the experimental results \cite{HouIJHMT2022}.

\begin{figure}[H]
	\centering
	\includegraphics[width=0.43\textwidth]{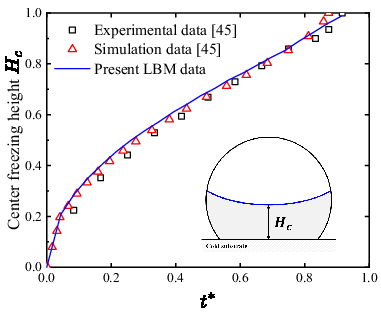}
	\caption{A comparison of the dimensionless center freezing front height $H_c$.}
	\label{fig5}
\end{figure}

To conduct a quantitative comparison between the present results and some available data, the time evolution of the square of the dimensionless center freezing height $H_c$ is shown in the Fig. \ref{fig5}, in which the results in Ref. \cite{HouIJHMT2022} are also incorporated. As seen from this figure, the evolution of the freezing front inside the droplet predicted by the LB method has a good agreement with the previous results \cite{HouIJHMT2022}, which also indicates that the phase-field base LB method can be used to predict the freezing process of the water droplet on the cold surface.

\begin{figure}[H]
	\centering
	\includegraphics[width=0.4\textwidth]{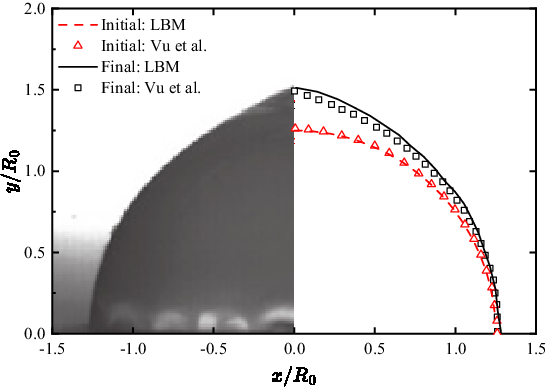}
	\caption{A comparison of the freezing profile among the present results, the referenced numerical results \cite{VuIJMF2018} and the experimental results \cite{MarinPRL2014}.}
	\label{fig6}
\end{figure}

In contrast to the spherical cap shape of frozen droplet observed in the above experiment, the pointy frozen droplet on the cold substrate have also been widely reported. To further validate the ability of present LB method in capturing the tip formation during freezing process, we perform a comparison of  current numerical results with the experimental results of Marin et al. \cite{MarinPRL2014} and the numerical solution of Vu et al. \cite{VuIJMF2018}. In the experiments of Marin et al. \cite{MarinPRL2014}, a $4-8 \mu \mathrm{L}$ droplet with the contact angle $\theta=90^{\circ}$ is gently deposited on a cold plate with the constant temperature $T=-44.1^{\circ} \mathrm{C}$. Fig. \ref{fig6} shows the initial and final freezing profiles of droplet, and it can be seen that the present results are in good agreement with the available experimental \cite{MarinPRL2014} and numerical data \cite{VuIJMF2018}. In addition, the predicted values of aspect ratio $H/R$ by the phase-field base LB method and experiments are 1.18 and 1.16, respectively, $H$ is defined as the final height of the ice drop and $R$ the radius of the wetted surface area.

\subsubsection{Freezing characteristics of droplet for different solid-to-liquid density ratios and wettability}

\begin{figure}[H]
	\centering
	\subfigure[Volume expansion after freezing]
	{
		\includegraphics[width=0.45\textwidth]{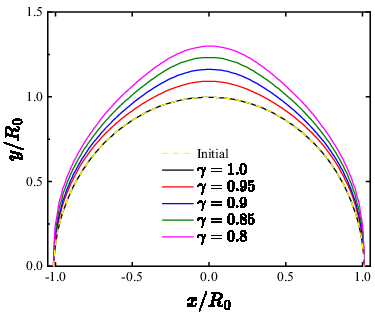} \label{fig7a}
	}
	\hspace{2mm}
	\subfigure[Volume shrinkage after freezing]
	{
		\includegraphics[width=0.45\textwidth]{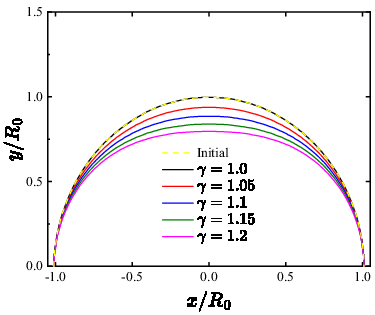} \label{fig7b}
	}	
	\caption{Effect of solid-to-liquid density ratio $\gamma = \rho_s/\rho_l$ on frozen droplet shapes at $\theta=90^{\circ}$, the dashed line represents the initial droplet profile.}
	\label{fig7}
	\end {figure}

The density difference between liquid and solid phases leads to volume changes of droplet during the freezing process. For example, the total volume of water droplet increases during freezing due to the decrease in average density. In the following, we intend to study the problem of droplet freezing on the cold surface at different solid-to-liquid density ratios. Fig. \ref{fig7} shows the final frozen droplet profiles under different solid-to-liquid density ratios $\gamma = \rho_s/\rho_l$. It is observed that compared to the initial liquid drop, the volume of droplet expanded for $\gamma \textless 1$, shrank for $\gamma \textgreater 1$, and remained unchanged for $\gamma = 1$. In addition, the expansion and shrinkage of water droplet mainly occurred in the y-direction, with little expansion or shrinkage in the radial direction, consistent with previous experiments \cite{SongEB2020}.

\begin{figure}[H]
	\centering
	\includegraphics[width=0.4\textwidth]{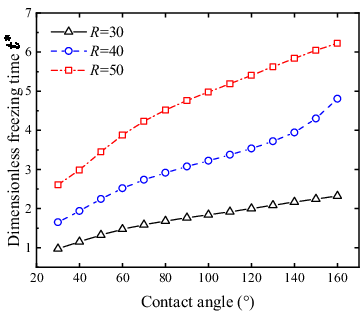} 
	\label{fig8a}
	\put(-210,175){\small (\textit{a})}
	\vspace{0.15cm}
	\includegraphics[width=0.4\textwidth]{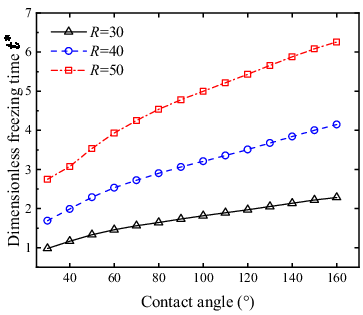} 
	\label{fig8b}
	\put(-210,175){\small (\textit{b})}
	
	\caption{Effect of contact angle on dimensionless freezing time $t^*$ of droplet for different initial radius $R$. (a) $\gamma = \rho_s/\rho_l <1$ ;(b) $\gamma = \rho_s/\rho_l > 1$}
	\label{fig8}
\end{figure}

To study the effect of surface wettability on the freezing process, we conducted simulations of droplet freezing on a cold surface with different contact angles varying from $\theta=30^{\circ}$ to $\theta=160^{\circ}$, while keeping the droplet volume and other parameters fixed. In our simulation, a droplet with radius $R$ is placed on a surface with contact angle $\theta$. Once the droplet reaches the prescribed contact angle $\theta$, a temperature field is applied. Fig. \ref{fig8} illustrates the effect of surface contact angle $\theta$ on the dimensionless freezing time $t^*$ for different initial droplet radii $R$ and solid-liquid density ratio $\gamma$. As shown in this figure, it is apparent that the freezing time increases with the increase of contact angle $\theta$ at the same initial liquid volume, and the increase of volume also leads to an increase in freezing time. The similar tendency has been reported by other researchers through numerical simulations \cite{ZhangIJHMT2018} and experiments \cite{ZhangIJTS2016, HuangETFS2012}.

\begin{figure}[H]
	\subfigure[$\theta=30^{\circ}$]{
		\includegraphics[width=0.3\textwidth]{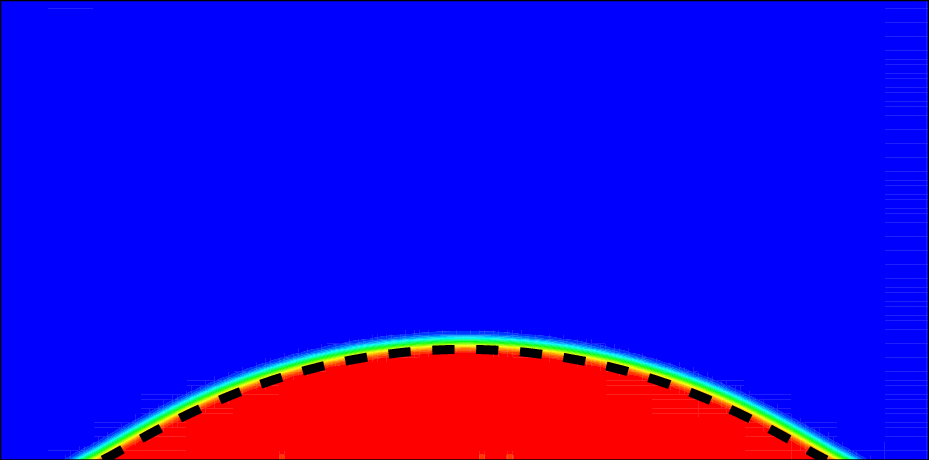} 
		\label{fig9a}
	}
	\hspace{2mm}
	\subfigure[$\theta=45^{\circ}$]{
		\includegraphics[width=0.3\textwidth]{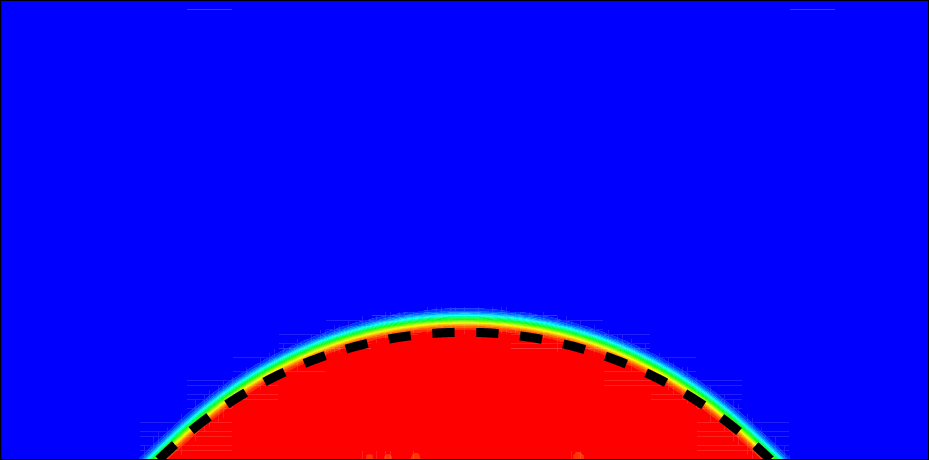} 
		\label{fig9b}
	}	
	\hspace{2mm}
	\subfigure[$\theta=60^{\circ}$]{
		\includegraphics[width=0.3\textwidth]{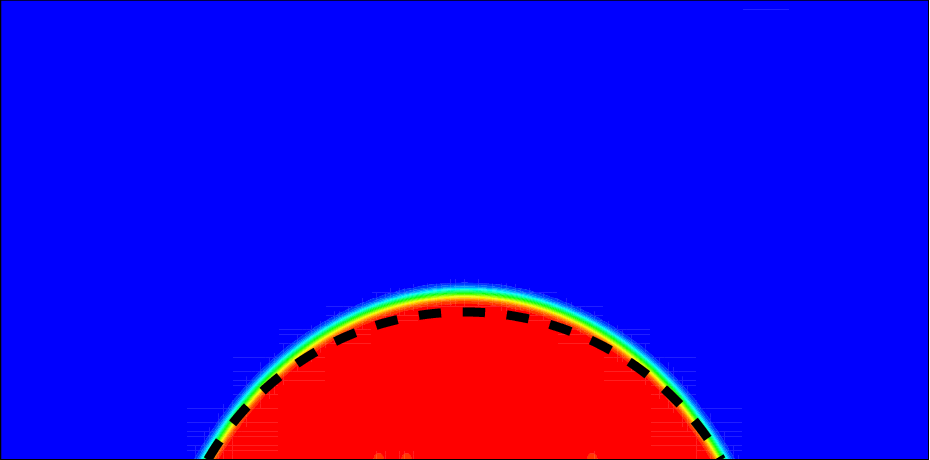} 
		\label{fig9c}
	}

	\subfigure[$\theta=90^{\circ}$]{
		\includegraphics[width=0.3\textwidth]{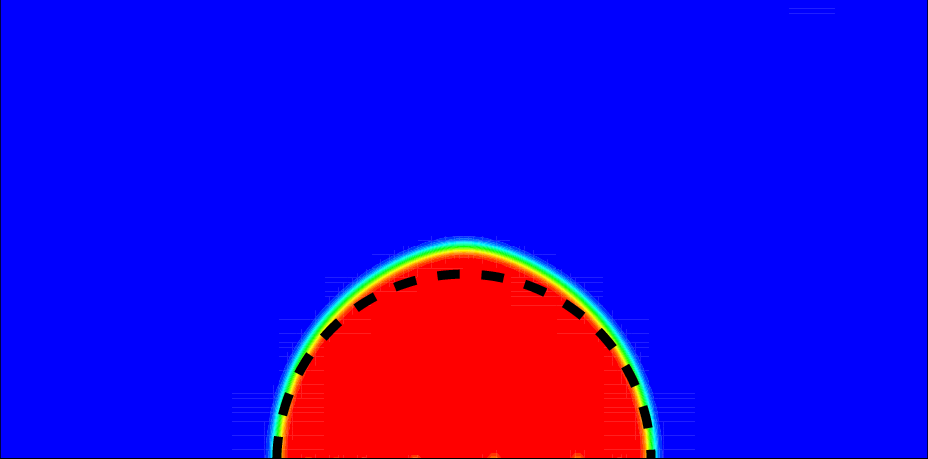} 
		\label{fig9d}
	}
	\hspace{2mm}
	\subfigure[$\theta=120^{\circ}$]{
		\includegraphics[width=0.3\textwidth]{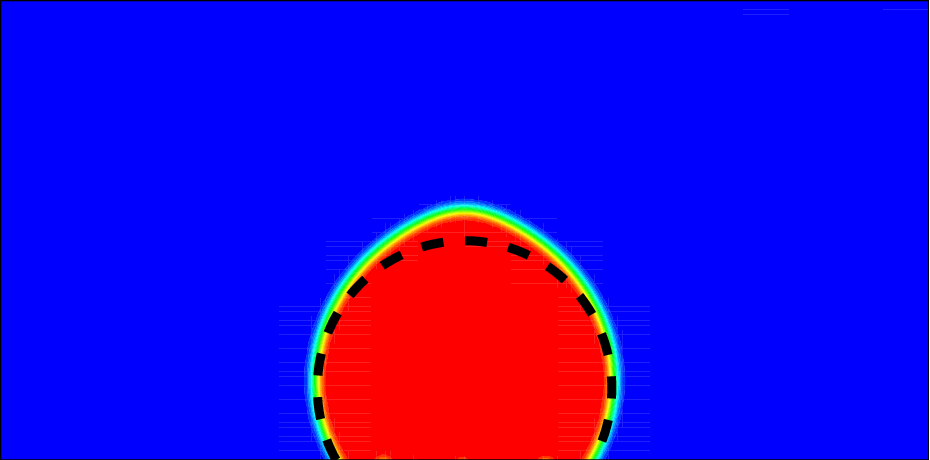} 
		\label{fig9e}
	}	
	\hspace{2mm}
	\subfigure[$\theta=150^{\circ}$]{
		\includegraphics[width=0.3\textwidth]{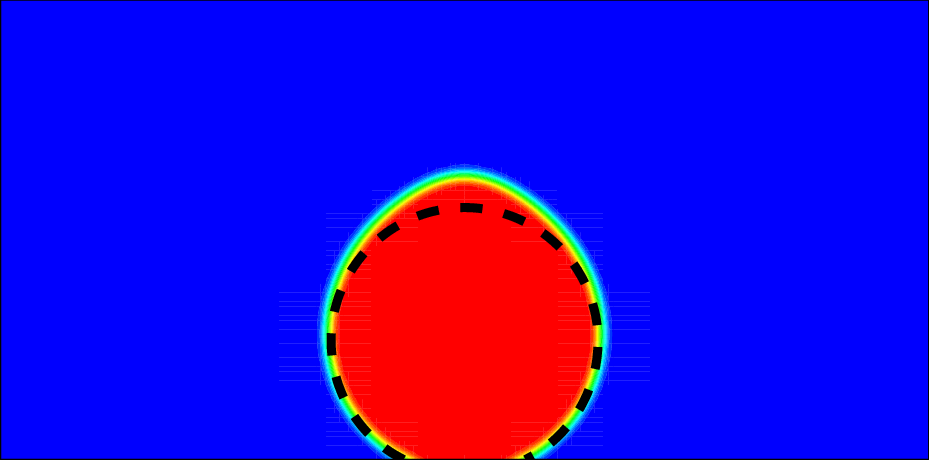} 
		\label{fig9f}
	}
	
	\caption{Final frozen droplet shape for different contact angles $\theta$ at $R=40$ for $\gamma=0.9<1$, the dash line represents the initial solidification front. Important simulation parameters are $\mathrm{Pr}=7.25$, $\mathrm{Ste}=0.25$, $C_{p,g}/C_{p,l}=1.0$ and $C_{p,g}/C_{p,s}=0.5$}
	\label{fig9}
\end {figure}

\begin{figure}[H]
	\subfigure[$\theta=30^{\circ}$]{
		\includegraphics[width=0.3\textwidth]{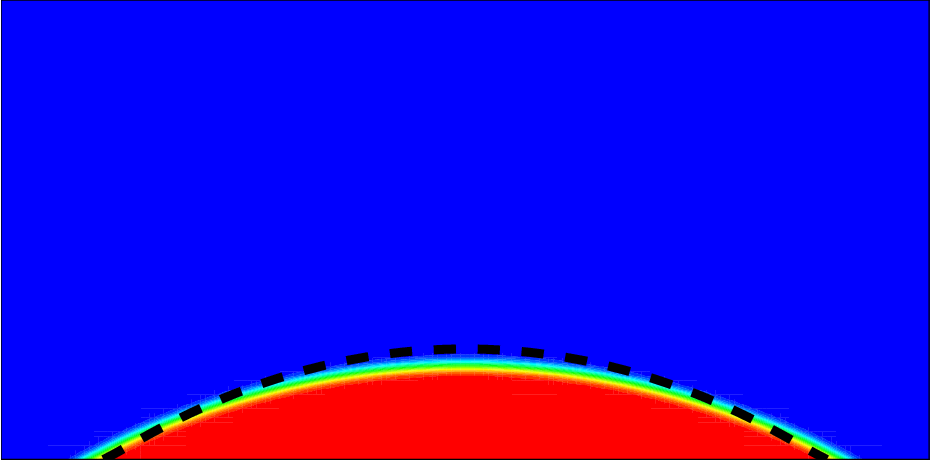} 
		\label{fig10a}
	}
	\hspace{2mm}
	\subfigure[$\theta=45^{\circ}$]{
		\includegraphics[width=0.3\textwidth]{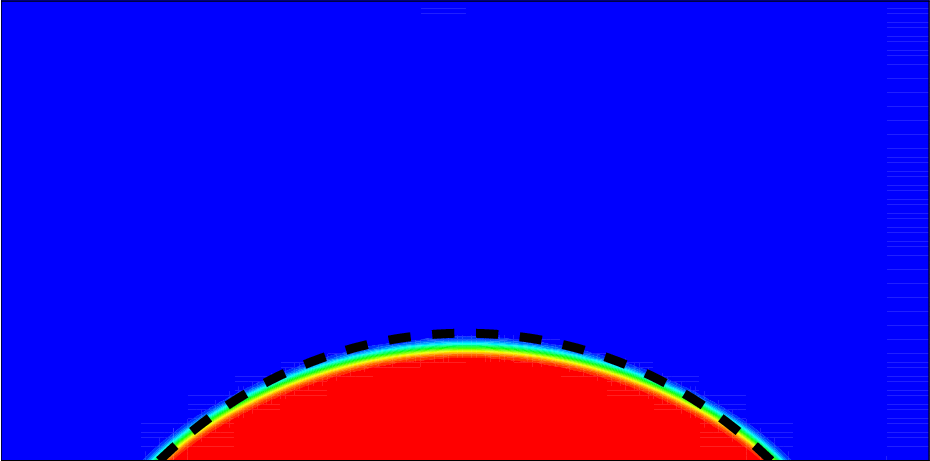} 
		\label{fig10b}
	}	
	\hspace{2mm}
	\subfigure[$\theta=60^{\circ}$]{
		\includegraphics[width=0.3\textwidth]{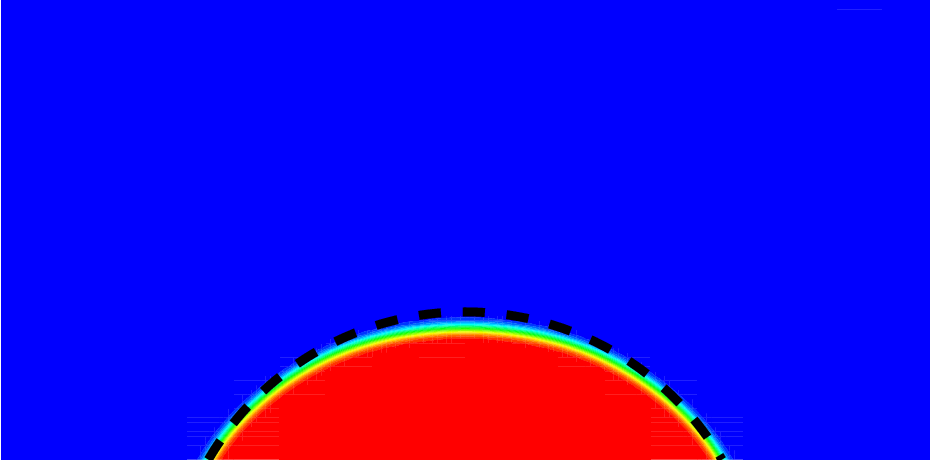} 
		\label{fig10c}
	}

	\subfigure[$\theta=90^{\circ}$]{
		\includegraphics[width=0.3\textwidth]{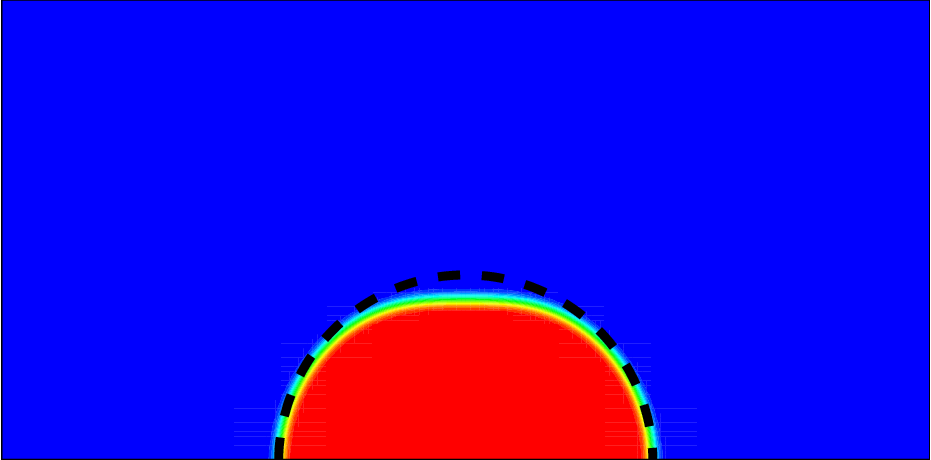} 
		\label{fig10d}
	}
	\hspace{2mm}
	\subfigure[$\theta=120^{\circ}$]{
		\includegraphics[width=0.3\textwidth]{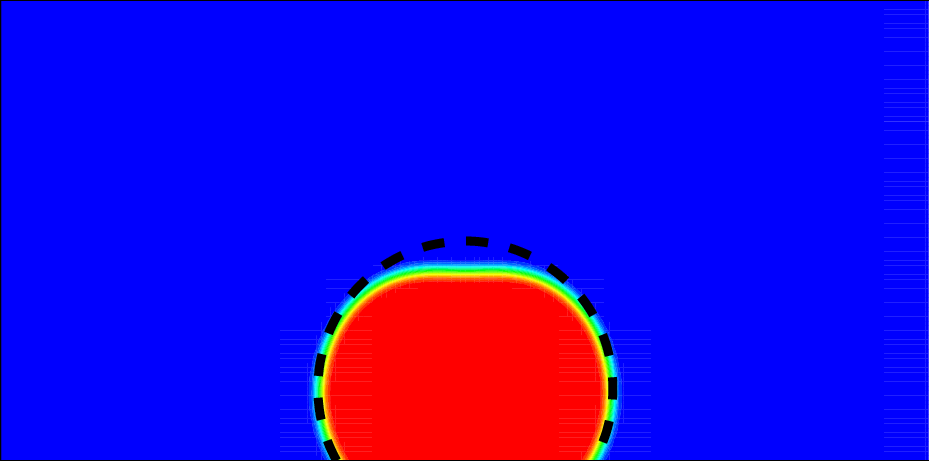} 
		\label{fig10e}
	}	
	\hspace{2mm}
	\subfigure[$\theta=150^{\circ}$]{
		\includegraphics[width=0.3\textwidth]{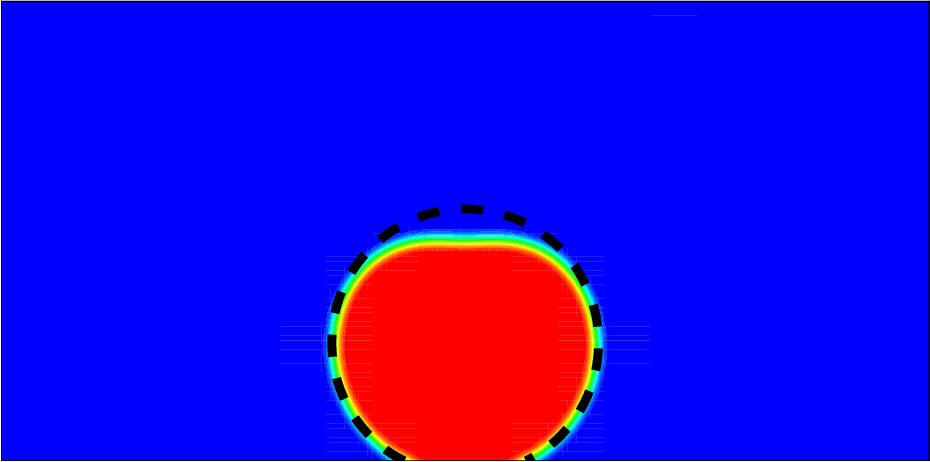} 
		\label{fig10f}
	}
	
	\caption{Final frozen droplet shape for different contact angles $\theta$ at $R=40$ for $\gamma=1.1>1$, the dash line represents the initial solidification front.}
	\label{fig10}
	\end {figure}

Fig. \ref{fig9} and Fig. \ref{fig10} show the final profile of frozen droplet at different contact angle for $\gamma<1$ and $\gamma>1$, respectively, and the dashed line represents the initial droplet with prescribed contact angle $\theta$. It can be seen that for $\gamma<1$, the gas-liquid interface gradually expands outward and eventually forms a conical tip at the top of the droplet for a larger contact angle. In contrast, the liquid-gas interface gradually shrinks inward and shrinks in volume for $\gamma>1$, resulting in the formation of a distinct plateau at the top of the droplet. In addition, a larger contact angle leads to a smaller initial base radius and a larger droplet initial height. The former reduces the contact area between the droplet and the cold surface, while the latter increases the thermal resistance. The combined effect of these factors slow down the release of latent heat and the freezing process. Therefore, a larger contact angle results in a lower thermal resistance, which delays the freezing process.

In addition, the rate of freezing in this isothermal freezing stage is mainly controlled by the rate of heat transfer from the substrate to the droplet and the rate of dissipation to the surrounding environment via convection. Considering that the droplet is generally very small in size, the natural convection between it and the air is practically weak and convective heat transfer is neglected. Therefore, the freezing rate is then mainly controlled by the rate of heat transfer between the droplet and the solid surface, which is described by the following equation \cite{OberliACIS2014}
\begin{equation}
	dQ/dt=-h_cL(T_w-T_0),
	\label{eqCon}
\end{equation}
where $ dQ/dt $ is the heat transfer rate ($\mathrm{J/s}$), $h_c$ is the heat transfer constant $(\mathrm{J/m^2~s~K})$, $L$ is the contact length $(\mathrm{m})$. According the Eq. (\ref{eqCon}), the heat transfer by means of conduction is proportional to the contact area or length. For the same volume of droplet, the contact area of surfaces with small contact angles is smaller than that of surfaces with large contact angles. Therefore, the thermal conductivity decreases with increasing contact angle, leading to a longer freezing time for large contact angle. In summary, we conducted a series of numerical simulations to study the freezing process of a droplet on a cold wall. The results showed that the model can accurately predict the freezing process and its influencing factors, and is in good agreement with previous experimental and analytical results. This suggests that the present LB model is suitable for simulating the freezing process on cold surfaces.

\subsection{Rising bubbles with solidification}
The evolution of gas bubbles during solidification process is widely encountered in many engineering applications, such as metalworking, pharmaceutical manufacturing and materials science \cite{SunIJHMT2016, LeeJIM2001}. Simulation of the solid–liquid–gas interaction during solidification process is a challenge due to the complex interfacial dynamics, the bubble deformation, and the heat transfer across the solid-liquid interface. Recently, Huang et al. \cite{HuangJCP2022} used a consistent and conservative phase-field model to simulate the solidification behavior of a liquid column with three circular gas bubbles, and their simulation results showed that the two bubbles at the lower position are eventually captured and frozen as two hollows, while the bubble at the upper position rises upwards under the action of buoyancy forces until it broken and merged with the gas-liquid interface. In this part, to further demonstrate the applicability of our LB method for solidification processes with gas bubbles, we will simulate the freezing process of liquid column containing bubbles.

The simulations are conducted on a uniform computational mesh with $Lx \times Ly=4l \times 3l$. A liquid column with a height of $H=1.8l$ was initially located at the bottom wall, and the remaining region was occupied by gas [can be seen in Fig. (\ref{fig10a})]. To be smooth across the interface, the initial order distribution can be set by Eq. (\ref{stefan}). The liquid column contains three circular gas bubbles with radii of $0.2l$, $0.24l$, and $0.3l$ from left to right, and their centers are located at $(l, 0.3l)$, $(2l, 1l)$, and $(3l, 0.4l)$, respectively. And the initial temperature of the system is set to $T_0$ and a low temperature $T_w$ is applied at the bottom. The periodic boundary conditions is conducted in the horizontal direction, the bottom and top planes are two solid walls imposed by the non-slip bounce back boundary condition. In addition, all the physical parameters used in this problem are the same as in the previous subsection, $\mathrm{Pr}=7.25$, $\mathrm{Ste}=0.02$, $k_{s}/k_{l}=3.8$, $C_{p,s}/C_{p,l}=0.5$ and $\rho_{s}/ \rho_{l}=0.9$. Furthermore, the surface tension force in Eq. (\ref{eq10}) is modified to $\mathbf{F}_s=f_s \mu_\phi \nabla \phi$, which ensures that the surface tension force only acts at the interface between the bubble and the liquid phase. With the assumption of the Boussinesq approximation, the the buoyancy force can be described as $\mathbf{F}_g=-\rho_0\mathbf{g} \beta\left(T-T_{ref}\right) $, where $\mathbf{g}$, $\beta$ and $T_{ref}$ are the acceleration of gravity, volumetric expansion coefficient and reference temperature, respectively.

Fig. (\ref{fig10}) illustrates the solidification process of liquid pool with gas bubbles, in which the solid, liquid and gas phases are filled with the white, blue, and orange colors, respectively. In the initial stage, the solidified phase begins to develop near the cold substrates, and the freezing front gradually moves upward. As the solidified phase grows, the two bubbles located at lower positions gradually freeze and form two hollows. Due to the much lower thermal conductivity of the gas phase than that of the liquid phase, the solidification is slower right above the two trapped gas bubbles, resulting in a V-shaped solid-liquid interface [can be seen in Fig. (\ref{fig10i})]. At the same time, the gas bubbles at the upper position gradually rise to the liquid-gas interface under the action of buoyancy and eventually rupture. Finally, the liquid pool completely solidifies, with two hollows on the left and right formed by the gas bubbles. It can be found that the present simulation results are in general agreement with those of Huang et al. \cite{HuangJCP2022}. This indicates that the present LB method can accurately capture the evolution of gas bubbles during solidification process.

\begin{figure}[H]
	\subfigure[$Fo=0$]{
		\includegraphics[width=0.3\textwidth]{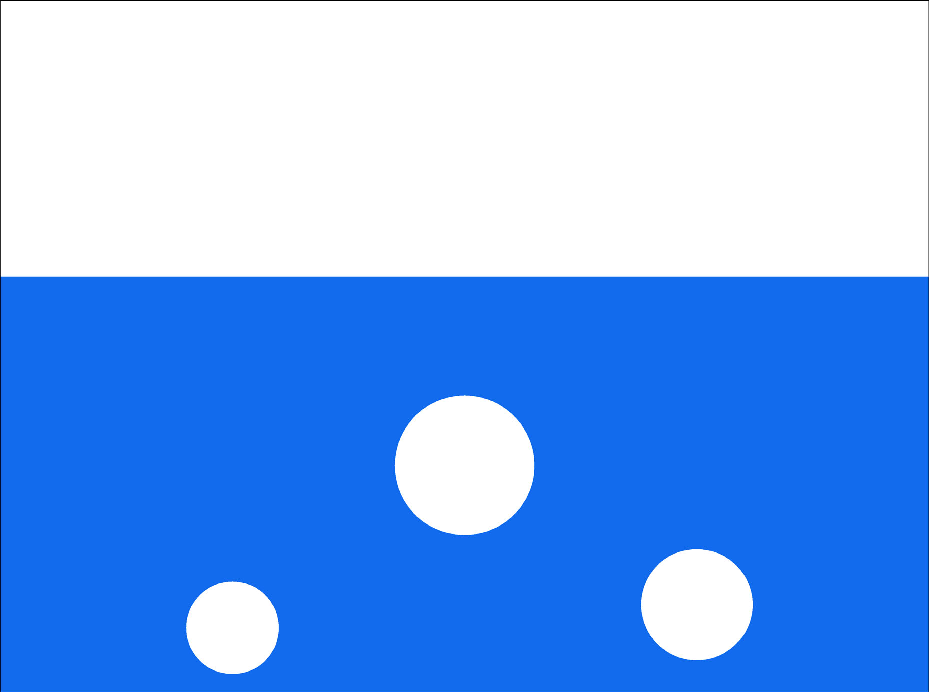} 
		\label{fig11a}
	}
	\hspace{2mm}
	\subfigure[$Fo=0.05$]{
		\includegraphics[width=0.3\textwidth]{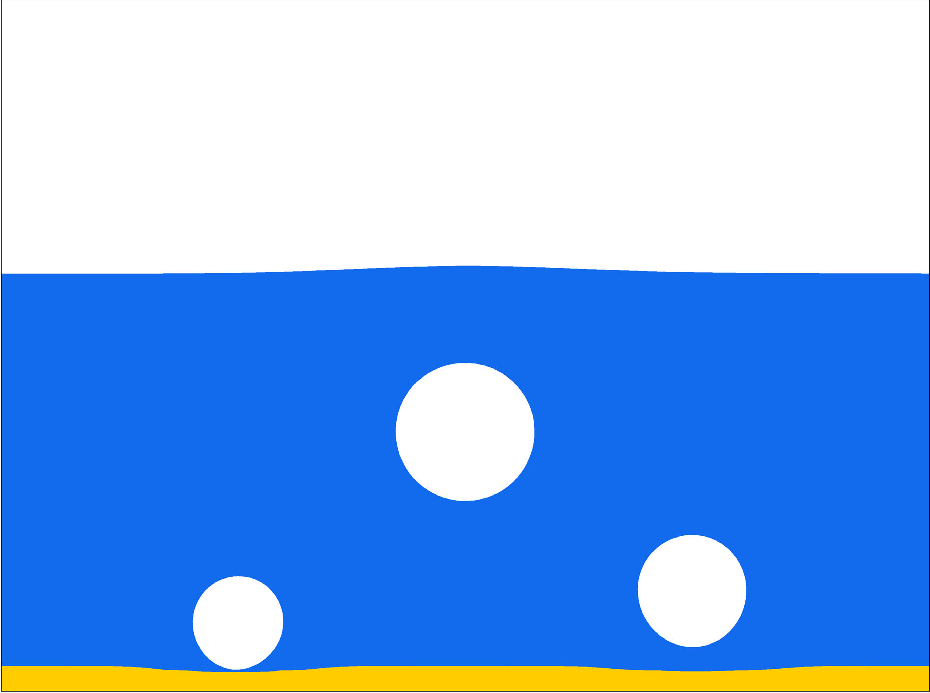} 
		\label{fig11b}
	}	
	\hspace{2mm}
	\subfigure[$Fo=0.1$]{
		\includegraphics[width=0.3\textwidth]{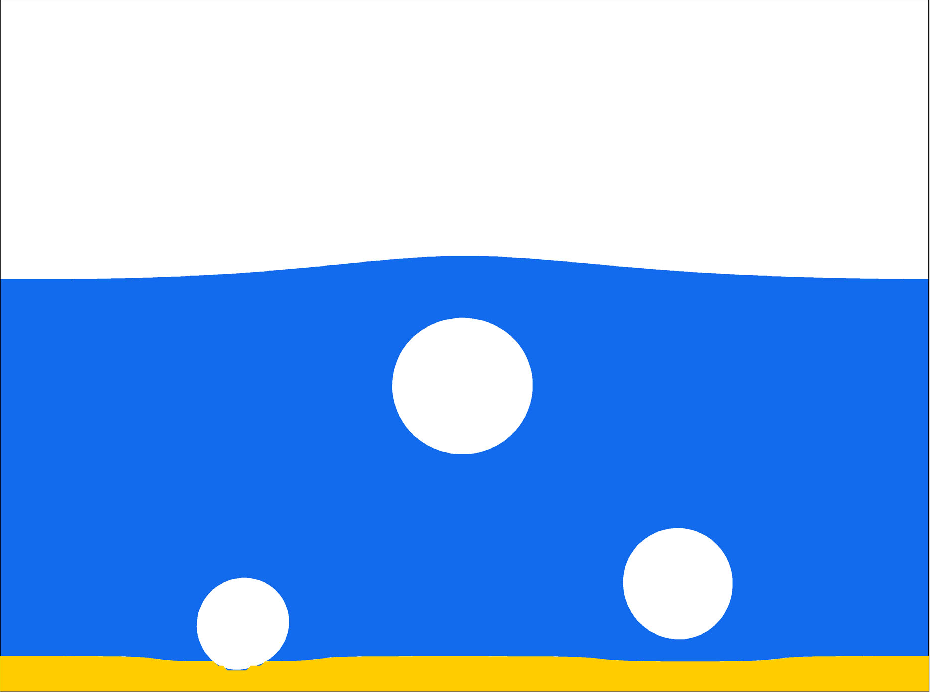} 
		\label{fig11c}
	}

	\subfigure[$Fo=0.2$]{
		\includegraphics[width=0.3\textwidth]{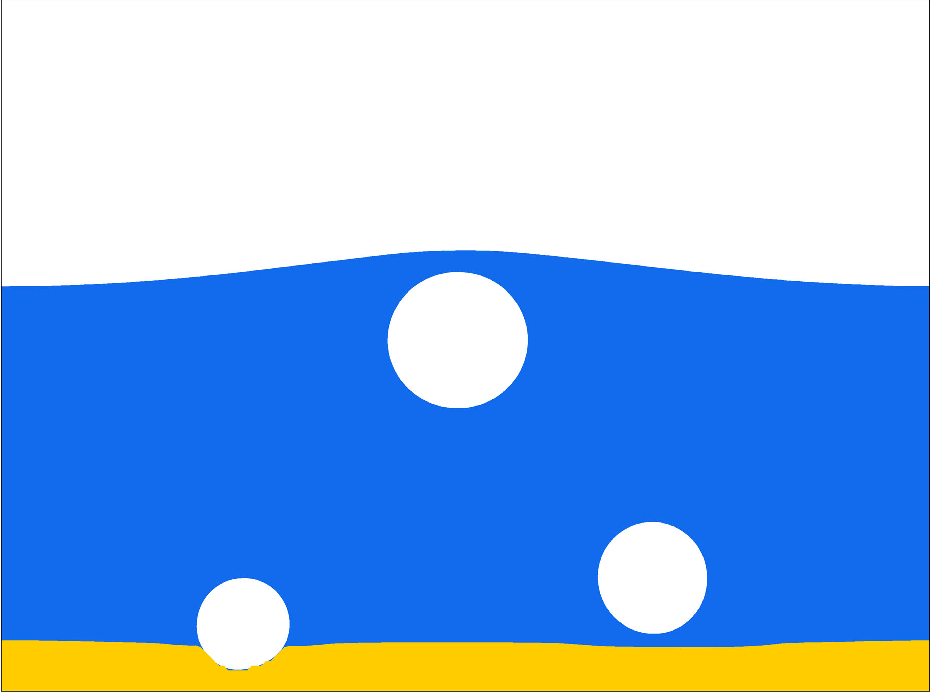} 
		\label{fig11d}
	}
	\hspace{2mm}
	\subfigure[$Fo=0.26$]{
		\includegraphics[width=0.3\textwidth]{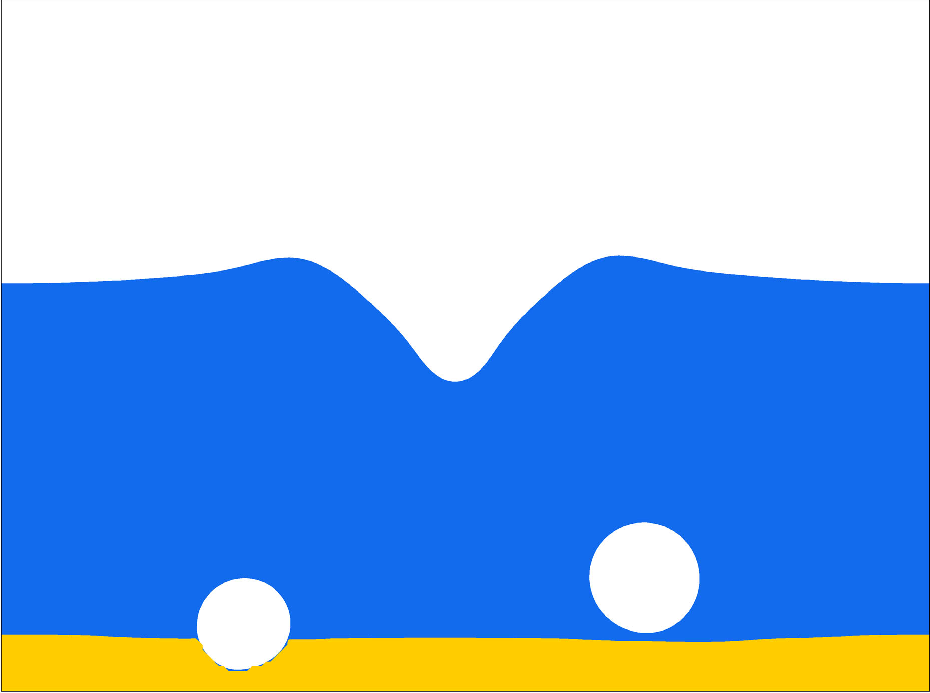} 
		\label{fig11e}
	}	
	\hspace{2mm}
	\subfigure[$Fo=0.31$]{
		\includegraphics[width=0.3\textwidth]{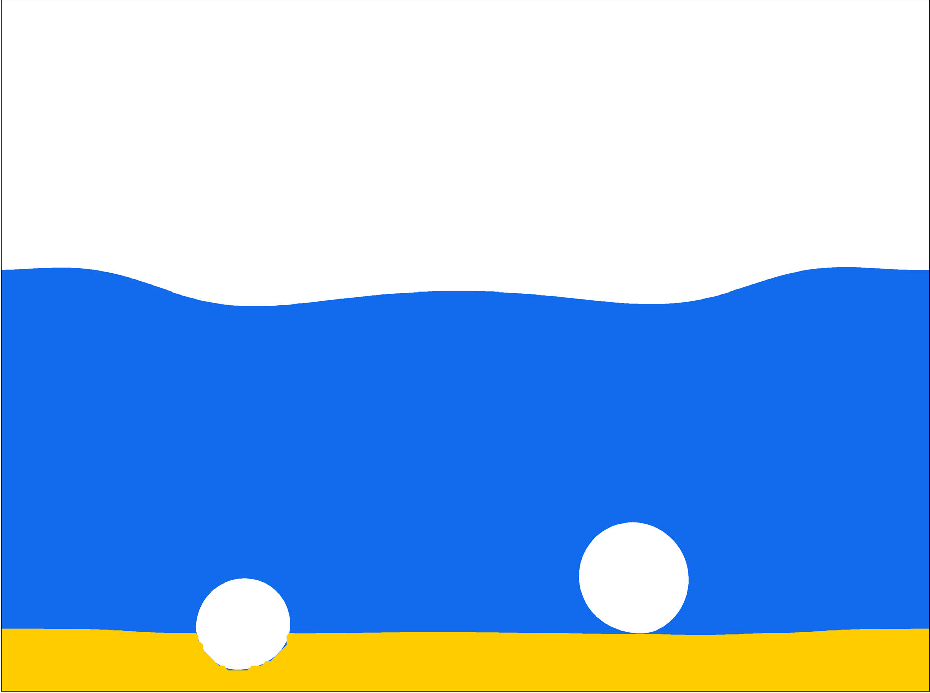} 
		\label{fig11f}
	}

	\subfigure[$Fo=0.41$]{
		\includegraphics[width=0.3\textwidth]{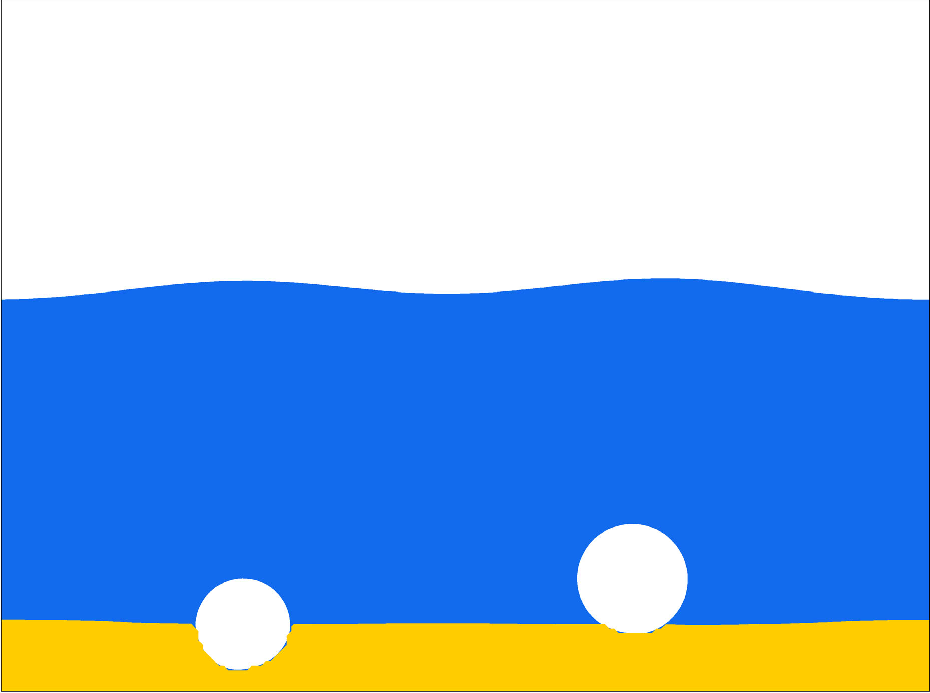} 
		\label{fig11j}
	}
	\hspace{2mm}
	\subfigure[$Fo=1.5$]{
		\includegraphics[width=0.3\textwidth]{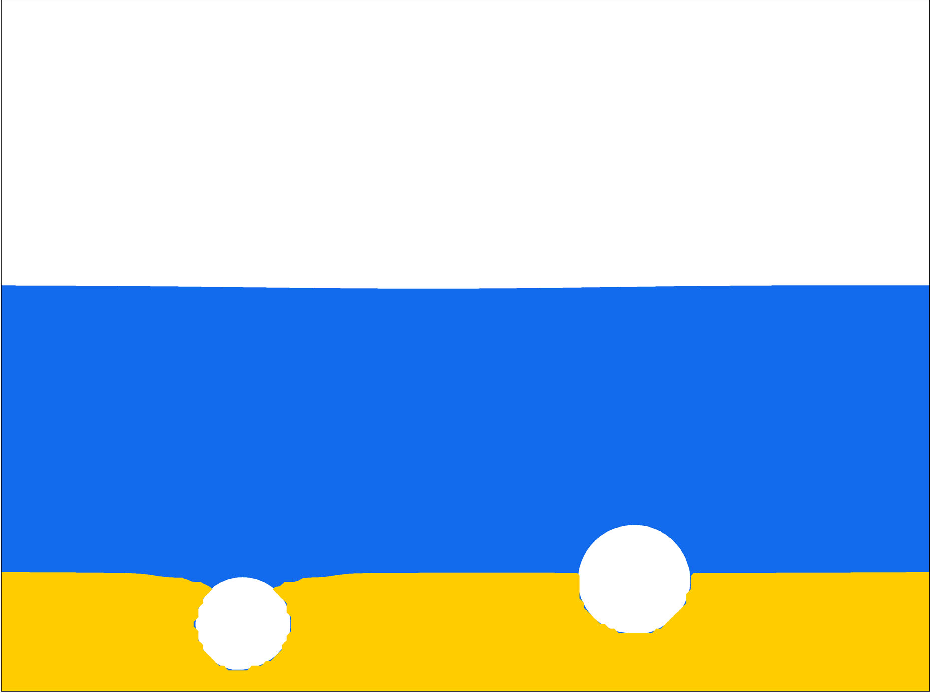} 
		\label{fig11k}
	}	
	\hspace{2mm}
	\subfigure[$Fo=4.2$]{
		\includegraphics[width=0.3\textwidth]{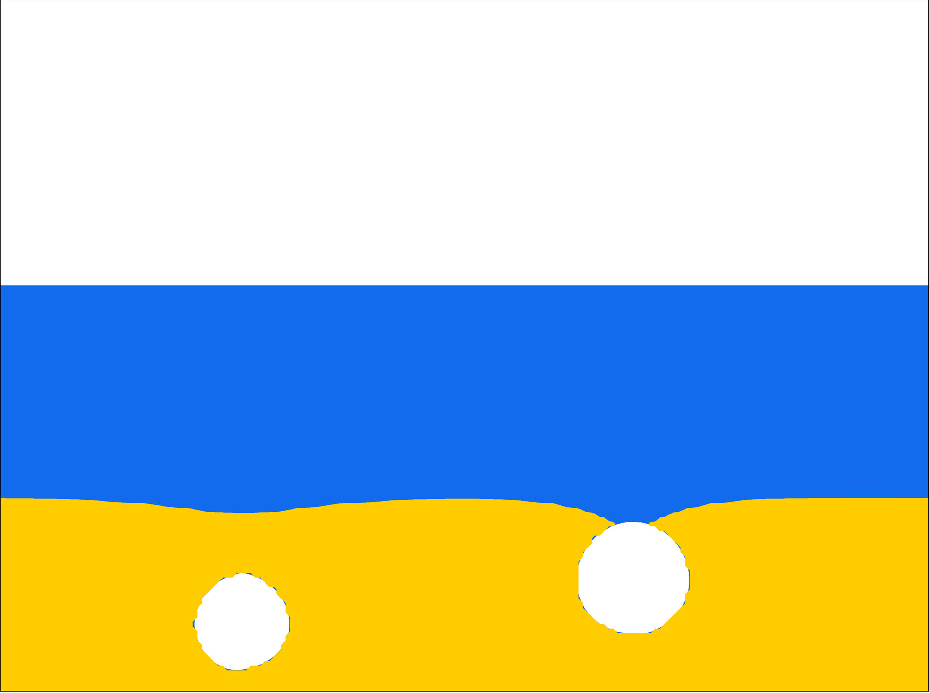} 
		\label{fig11l}
	}

	\subfigure[$Fo=5.2$]{
		\includegraphics[width=0.3\textwidth]{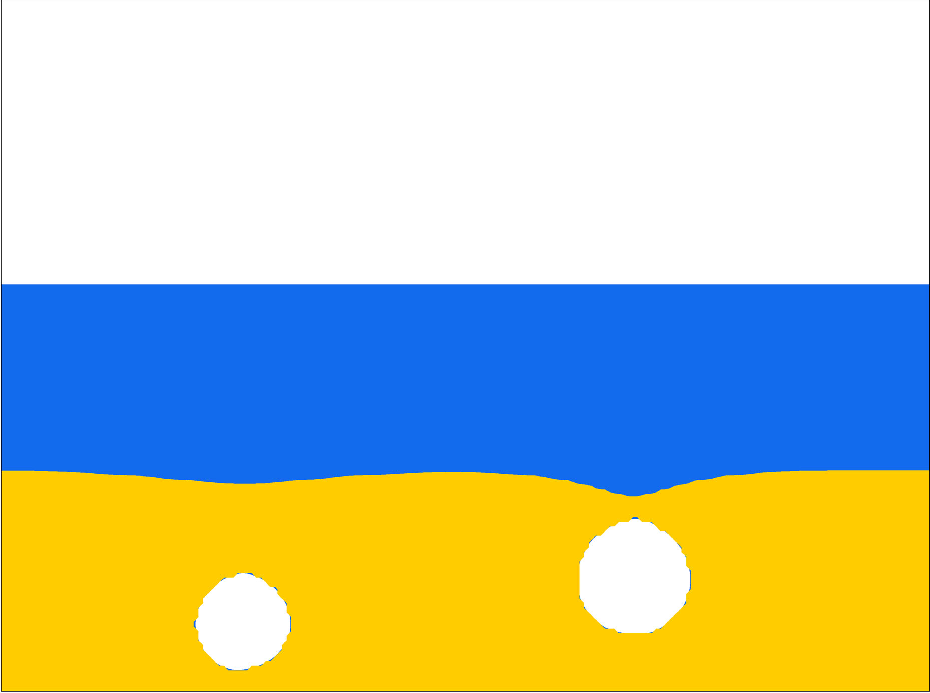} 
		\label{fig11m}
	}
	\hspace{2mm}
	\subfigure[$Fo=10.4$]{
		\includegraphics[width=0.3\textwidth]{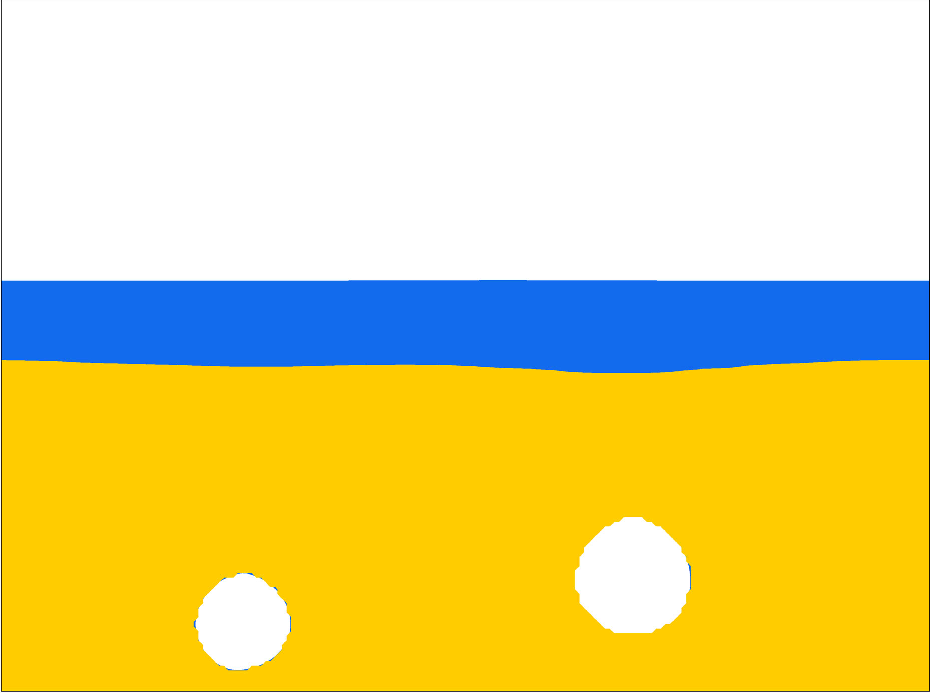} 
		\label{fig11n}
	}	
	\hspace{2mm}
	\subfigure[$Fo=15.6$]{
		\includegraphics[width=0.3\textwidth]{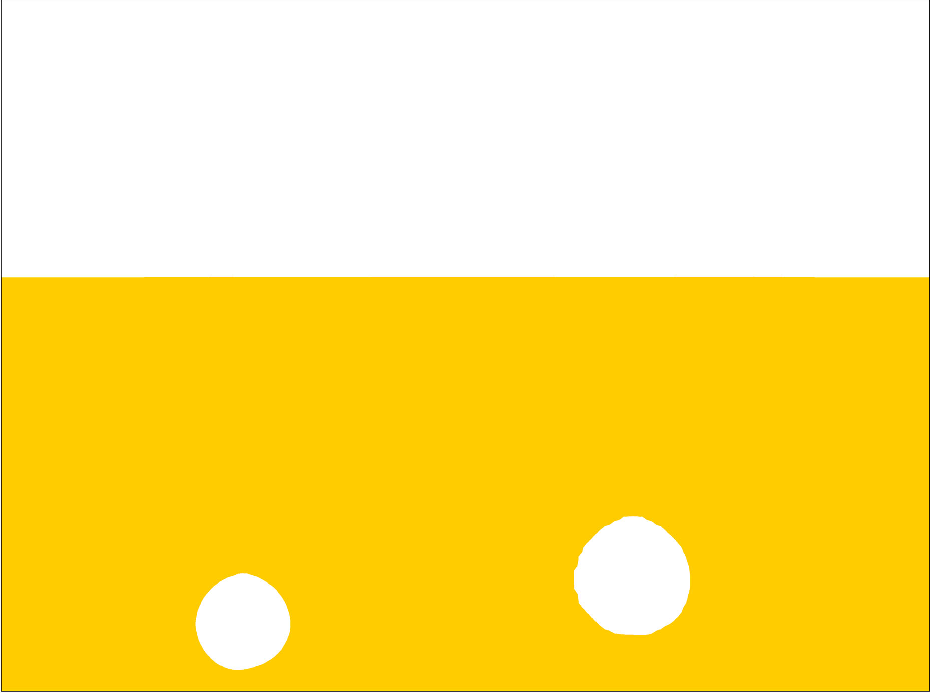} 
		\label{fig11o}
	}
	
	\caption{Results of rising bubbles with solidification. The gas, liquid, and solid phases are filled by the white, blue and orange colors, respectively.}
	\label{fig11}
	\end {figure}

\section{Conclusions}
\label{sec5}

In this work, we propose a phase-field LB method to simulate the freezing process with volume change in a gas-liquid-solid system where the volume expansion or shrinkage caused by the density difference between liquid and solid is considered through adding a source term in the continuity equation. The model is first validated by simulating the solidification of a liquid column, and the results show that the proposed LB method is able to study the liquid-solid phase change with the volume change. Then, the LB method is applied to investigate the freezing dynamics of droplet on a cold substrate, and it is found that the present numerical results are in good agreement with the experimental data. Furthermore, we focus on the effects of several key parameters on the freezing of sessile droplet on a cold surface, such as the ratio of solid density to liquid density, the contact angle and the volume of droplet, and find that the solidification time increases with the increase of contact angle and droplet volume, which is consistent with previous experimental results. In addition, the solid-liquid density ratio has a significant influence on the evolution of the droplet shape. For the case with volume expansion $(\rho_{s}  \textless \rho_{l})$, the frozen droplet tends to form a conical shape on the upper surface, while for the case with volume shrinkage $(\rho_{s} \textgreater \rho_{l})$, a distinctive plateau is formed at the top of the frozen droplet. Finally, a challenging problem of liquid column solidification with bubbles is also considered, and the rising and deformation behavior of bubbles during solidification can be captured. In summary, the present LB method is effective and accurate in the study of the solidification problems.

\section*{Acknowledgments}
This work was financially supported by the National Natural Science Foundation of China (Grants No. 12072127 and No. 51836003), the Interdisciplinary Research Program of Hust (2023JCJY002), and the Fundamental Research Funds for the Central Universities, Hust (No. 2023JYCXJJ046). The computation is completed in the HPC Platform of Huazhong University of Science and Technology.


\end{document}